\documentclass[]{aa}  

\DeclareRobustCommand{\VAN}[3]{#2}
\let\VANthebibliography\thebibliography
\def\thebibliography{\DeclareRobustCommand{\VAN}[3]{##3}\VANthebibliography}

\usepackage[colorlinks=true,
    linkcolor=blue, citecolor=blue, filecolor=blue, urlcolor=blue]{hyperref}
\usepackage{graphicx}	
\usepackage{amsmath}	
\usepackage{amssymb}	
\usepackage{array}
\usepackage{multirow}
\usepackage{bigdelim}
\usepackage[varg]{txfonts}
\usepackage{xcolor}
\usepackage[version=4]{mhchem}
\usepackage{cuted}

\begin{document}

\title{Resolving dense photodissociation regions: the structure of photochemical fronts in three-dimensional gas distributions}
\titlerunning{Resolving dense PDRs}
\authorrunning{Gaches et al.}

\author{Brandt A. L. Gaches
        \inst{1}\corrauth{E-mail: brandt.gaches@uni-due.de}
        \and
        Thomas G. Bisbas\inst{2}\email{tbisbas@zhejianglab.com}
        \and
        Lothar Brendel\inst{1}\email{lothar.brendel@uni-due.de}
        \and
        Zhengping Zhu\inst{2}\email{zhuzhp@zhejianglab.org}
        }
\institute{
Faculty of Physics, University of Duisburg-Essen, Lotharstraße 1, 47057 Duisburg, Germany \and
Research Center for Computational Earth and Space Science, Zhejiang Lab No. 2880 Wenyi West Road, Yuhang District, Hangzhou, Zhejiang Province, 311100, CN
}

\date{Accepted XXX. Received YYY; in original form ZZZ}

\abstract{For decades, the Orion Bar has been the prototypical photodissociation region (PDR). Viewed nearly edge-on, it offers a unique window into the stratified chemical structure of the atomic-to-molecular transition of the interstellar medium. Understanding its photochemistry is essential to interpreting key observations originating from dense PDRs.}{ALMA and JWST observations reveal that \ce{H2} photodissociation front overlaps with the \ce{C+} recombination front at their angular resolutions and exhibits a complex spatial morphology. Despite considerable theoretical effort, existing modeling approaches based on simplified geometrical assumptions have difficulties reproducing the spatial emission structure. Our aim is to investigate the response of photochemistry in realistic three-dimensional density distributions, using the Orion Bar as a representative application.}{We present the first fully three-dimensional high-resolution model of an Orion Bar analogue that resolves the relevant photochemical fronts using the upgraded steady-state {\sc 3d-pdr} PDR code, which allows for the treatment of plane-irradiatation and for the solution of the non-LTE \ce{H2} rovibrational levels.}{We find that the \ce{H2} dissociation front is characterized by a complex surface that overlaps with the \ce{C+} recombination front. Our 3D model can reproduce the complex morphology of \ce{H2} emission seen in observations, in particular the arc- and filament-like features, and provides a physical explanation of its chemistry.}{The overlapping \ce{H2} dissociation and \ce{C+} recombination fronts and the spatial emission morphology can be explained due to the three-dimensional gas distribution, resulting in shadowing and shielding by dense substructures. Our results mark a turning point for astrochemistry, where three-dimensional steady-state models can deliver fundamentally new insights into the chemistry and fundamental structure of the interstellar medium.}


\maketitle
\nolinenumbers

\section{Introduction}\label{sec:intro}
Photodissociation regions (PDRs), sometimes called photon-dominated regions, are regions where the heating and chemistry are dominated by far-ultraviolet (FUV, $6<h\nu<13.6\,{\rm eV}$) radiation and are a fundamental component of the interstellar medium (ISM) \citep{Hollenbach1997, Wolfire2022}. Classical PDRs exist at the interface between the atomic ISM and dense molecular clouds, where the FUV radiation of massive stars drives an intense photochemistry. The surfaces of these PDRs, characterized by the transition from atomic (H) to molecular hydrogen (\ce{H2}), exhibit warm gas and dust temperatures that enable an interesting and complex photochemistry. These regions play an important role in our understanding of the molecular chemistry of the ISM. Most of the neutral gas mass in galaxies exists in PDRs \citep{Wolfire2022} and much of the diagnostic line emission from galaxies used to understand the formation of stars and evolution of the ISM originates from them. Therefore, accurate and high fidelity models of the heating, cooling, and chemistry within PDRs are essential for our understanding of the lifecycle of gas in galaxies. Canonical models of PDRs and UV-excitation of molecular hydrogen date back four decades \citep{Tielens1985,vanDishoeck1986, Black1987, Sternberg1989a,Sternberg1989b,Bertoldi1996, Draine1996, Hollenbach1997,LePetit2006}.

The Orion Bar has been used as a prototypical PDR since it is, by cosmic coincidence, observed to be nearly edge-on \citep{Tielens1993, Hogerheijde1995, Walmsley2000, Goicoechea2016, Habart2024}. While the edge-on nature makes constraining the dynamics more difficult, it enables an unprecedented and exemplary view of the photochemistry. The Orion Bar is close to us, 414 pc \citep{Menten2007}, with a well-constrained radiation field, irradiated primarily by the O7-type star $\theta^1$ Ori C in the Orion Nebula.

Due to its convenient geometry, proximity, and well-prescribed radiation field, the Orion Bar has become a standard benchmark for PDR models which are essential in interpreting observations of many molecules and neutral atoms in the ISM \citep{Roellig2007}. In the past decade, there have been significant observations that have challenged the status-quo of the current PDR chemical models: high-resolution observations from the Atacama Large Millimeter Array (ALMA) of \ce{HCO+} (J = 4-3) and \ce{CO} (J = 3-2) \citep{Goicoechea2016}, the PDRs4All program with the James Webb Space Telescope (JWST) Program \citep{Berne2022}, which observed the Orion Bar to high spatial and spectral resolution in the near infrared (NIR) and mid-infrared (MIR) \citep{Habart2024,Peeters2024}, and ground-based Keck/NIRC2 near-infrared observations \citep{Habart2023}. These observations showed that the transition region between H/\ce{H2} (hereafter H-front) and the transition between \ce{C+}/C/CO (hereafter C-front) appear to overlap \citep{Goicoechea2016, Habart2024, Goicoechea2025}. This also agrees with previous observations of radio recombination lines that found C recombination emission sometimes even preceding the \ce{H2} dissociation front \citep{Wyrowski1997, Cuadrado2019}. High-pressure isobaric models with the sophisticated one-dimensional PDR model Meudon \citep{LePetit2006} and one-dimensional chemodynamical models \citep{Bron2018, Kirsanova2019, Pomelnikov2025} can reproduce the closeness of the fronts. However, they can have difficulties matching the overall line intensities and, due to their one-dimensional nature, cannot model multi-dimensional spatial features.

The observations have also revealed the highly complex morphology of the PDR front, with many crenulations, voids, and an apparent tiered structure. The PDRs4All NIRCAM observations, for instance, revealed three different dissociation fronts within their field of view attributed to a tiered structure \citep{Habart2024}. NIR line emission and rotational molecular emission observed with ALMA shows a highly structured gas behind the gas, characterized by a dense clumpy structure embedded in moderate density gas. When interpreting these fronts with one dimensional slab models, inclination effects must be accounted for by fitting with geometric correction terms. Even then, multiple dissociation fronts along a particular cut of the data can confuse and complicate analysis. Observations have demonstrated that these PDR surfaces do not exhibit clean one-dimensional morphologies: three dimensional models are uniquely capable and necessary to investigate the morphology and photochemistry of the surfaces. It is worth highlighting the interesting and informative work using {\sc Kosma-$\tau$} \citep{Rollig2022} that reconstructed the 3D morphology of the Orion Bar using a voxel-based (3D pixel) statistical combination of 1D clumpy models \citep{Andree-Labsch2017}. However, as a statistical combination of one-dimensional models, it was not able to fully resolve the structure of the \ce{H2} photodissociation front and lacked a solution of the \ce{H2} rovibrational levels for the NIR \ce{H2} emission.

We present here a novel high-resolution three-dimensional PDR model using the public PDR code {\sc 3d-pdr} \citep{Bisbas2012}, upgraded to include the {\sc raytheia} ray-tracing algorithm \citet{Zhu26} and a new numerical solver for the rovibrational level populations of \ce{H2}. We use a fractal morphology of the gas density as a proxy for the PDR region to reproduce the distribution of gas induced by turbulent motions in the natal molecular cloud. This allows us to vary the fractal dimension, $\mathcal{D}$, which changes how clumpy the gas distribution is, thus varying the size and nature of the dense substructures. We find that breaking the assumption of a one-dimensional slab morphology to simulate a three-dimensional PDR allows the model to successfully capture the behavior of the overlapping, or even preceding, C- and H-fronts and provides important insights into the NIR \ce{H2} emission observed with JWST. Our model demonstrates that three-dimensionality is necessary to fully capture the relation between the photochemical fronts and comparing to spatially-resolved spectral-imaging observations of PDRs.

In Section \ref{sec:method} we present the our numerical methods. In particular, it describes the new plane-irradiation scheme and \ce{H2} rovibrational level population solver. In Section \ref{sec:res} we present the results of these first models, briefly describe the temperatures and chemistry, and explore the morphology and behavior of the photochemical fronts. Finally, in Section \ref{sec:disc} we discuss the implications of these results and present our main conclusions.

\section{Methods}\label{sec:method}
\subsection{Fractal PDR density model}
Observations of resolved dense PDRs have shown that they exhibit a complicated stratified structure as a result of turbulence and instabilities from the HII region cavity wall pushing into the surrounding medium or from a photoevaporating natal molecular cloud \citep{Goicoechea2016}. Fully self-consistently simulating the dynamics and chemistry of these regions, such that the various photochemical fronts of interest are resolved, would be significantly taxing in 3D. As a compromise, we use a fractal density distribution as a proxy to model the three-dimensional hydrodynamical structure. This was chosen to investigate the role of gas clumpiness and variance in the density distribution. Fractals have been found to reasonably well reproduce the gas morphologies and statistics of turbulent clouds, with fractal dimensions ranging from 2 to 3 depending in particular on the turbulence \citep{Scalo1990, Elmegreen1996, Beattie2019}. 

We first constructed the fractal for the whole domain in Fourier space, similar to the construction of fractal molecular clouds \citep{Stutzki1998, Walch2012}. In Fourier space, the field was constructed with random phases and with an amplitude according to a power-law, $P(k) = k^{-n}$, where $n = 2 \cdot(4 - \mathcal{D})$ and $\mathcal{D}$ is the fractal dimension. We use the Python package {\sc FyeldGenerator}\footnote{\url{https://github.com/cphyc/FyeldGenerator}} to construct the initial Gaussian field, $\mathcal{F}_0$. After generating it, the field is normalized as $\mathcal{F} = \mathcal{F}_0/\sigma(\mathcal{F}_0)$. Varying the fractal dimension allows control of the clumpiness of the density distribution. The fractal is centered on the highest density region. The Gaussian field is exponentiated and scaled for the desired log-normal density distribution, 
\begin{equation}
    \frac{n_{H,\mathrm{cloud}}}{\mathrm{cm}^{-3}} = e^{n' + \mathcal{F} \cdot\sigma},
\end{equation}
where $n' = \ln(n_0/\mathrm{cm}^{-3})$ is the log of the center of the density PDF, $n_0$, and $\sigma$ is the spread of the log normal. Once the full-domain fractal is generated, it is truncated by a softened Heaviside function to produce a uniform-density ``cavity'' which is filled with gas density of $n_H = 10^4$ cm$^{-3}$, resembling the density in the neutral and mostly atomic PDR layers. The end result is a distribution qualitatively similar to dense PDRs, with a lower-density component for the atomic PDR layer (here called the cavity) followed upstream by a highly structured dense gas distribution at PDR front, caused by, e.g., gas swept up in an expanding shell or from a surrounding photoevaporating natal molecular cloud.

We utilize a range of fractal dimensions, from $\mathcal{D} = 2.0 - 2.7$, ranging from large correlated structures to homogeneous distributions of small clumps. The choice of $\sigma = 1$ is appropriate for the observed non-thermal line width and gas temperatures seen from higher J level CO and \ce{HCO+} \citep{Goicoechea2016}. Low fractal dimensions, characterized by large, correlated density structures, would correspond to compressive motions, such as material being swept up along the Bar, or via highly supersonic turbulence. Conversely, high fractal dimensions exhibit more homogeneously distributed clumps, resulting from, e.g., turbulence with lower characteristic Mach numbers \citep[see e.g.,][]{Beattie2019}. Simulations of HII expansions into fractal molecular clouds have shown that low fractal dimension molecular clouds lead to shell-dominated features, while higher natal fractal dimensions lead to well-defined pillar structures \citep{Walch2012}. The chosen range of fractal dimensions is consistent with those found in interstellar clouds \citep{Elmegreen1996, Stutzki1998, Sanchez2007}.

\subsection{Chemical model}
We used the public PDR code, {\sc 3d-pdr}\footnote{\url{https://github.com/itamos-ism/3D-PDR}} \citep{Bisbas2012}, now upgraded with the new ray-tracing algorithm {\sc raytheia} \citep{Zhu26}, which solves the steady-state gas-phase chemistry, radiation transfer, non-LTE level populations, and heating and cooling. For this work, we utilized a subset of the UMIST network consisting of 108 species, primarily with 4 atoms or less, with larger molecules necessary for the chemistry: \ce{NH4+}, \ce{H3CO+}, \ce{C2H3+}, \ce{H3S2+}, and \ce{CH5+}. The species are connected via 1828 reactions. All reaction rates in this work have been updated to the most recent release of UMIST2022 \citep{Millar2024} and KIDA \citep{Wakelam2024}. The chemical network was developed as a reduced network to model the chemistry of important simple nitrogen-bearing species (\ce{HCN}, \ce{HNC}, \ce{N2H+}, and \ce{NH3}), simple hydrocarbons (\ce{CH}, \ce{CH_n^{(+)}}, \ce{C2H}), and sulfur-bearing species (\ce{S}, \ce{S+}, \ce{HS+}, \ce{H2S}). We included non-LTE level populations of \ce{C+}, C, O, and CO for line cooling. {\sc 3d-pdr} assumes that scattering is negligible and uses the Large Velocity Gradient (LVG) escape probability formalism \citep{Sobolev1960, deJong1975} to account for cooling due to line emission. More details can be found in \citet{Bisbas2012} and \citet{Zhu26}.

We included excited state chemistry induced by vibrationally excited \ce{H2}. For \ce{H2 + C+ -> CH+ +H} we use the state-specific reactions from \citet{Zanchet2013, Herraez-Aguilar2014} as implemented in the Meudon model. For \ce{H2 + S+ -> HS+ + H} we use the state-specific reactions from \citet{Zanchet2019, Goicoechea2021}. Other endothermic reactions which become possible via excited-state chemistry with \ce{H2}, are included under the assumption that the endothermicity of reactions with \ce{H2} is decreased by the energy level of a given ground state rovibrational level. We note that this is in general a very simple assumption in comparison to more accurate general methods such as those in \citet{Agundez2010, Zanchet2013, Herraez-Aguilar2014}. In this manuscript, we define rovibrationally-excited \ce{H2}, \ce{H2^*}, to be the \ce{H2} excited above 2.6 eV to match the canonical definition \citep{Tielens1985}. This is computed directly from the level population (see below). It is not used directly in the chemistry, but included as an output for analysis.

We implemented a new external radiation transfer mode in {\sc 3d-pdr} to enable one-sided illumination. The radiation transfer uses the {\sc Healpix} scheme for ray orientations, with a number of rays equal to $N_{\rm rays} = 12 \times 4^\ell$, where $\ell$ is the raytrace level of refinement controlling the angular resolution. We implemented the one-sided illumination scheme taking advantage that the line cooling and column densities for self-shielding use the same rays. The new mode allows users to specify the radiation source by domain boundary, which defines a direction vector $\hat{\chi}$ that points from each cell toward the designated face (e.g., for radiation originating from the $-x$ boundary, $\hat{\chi} = -\hat{x}$)

During the FUV radiation field calculation, we restrict the ray-tracing integration toward the irradiating surface by including only rays with direction $\hat{r}$ that satisfy $\hat{\chi} \cdot \hat{r} \ge 0$.
For rays that fulfill this requirement, we check to see if that ray will intersect with the domain boundary of interest. The effective surface radiation field for a ray, $j$, that satisfies these criteria, is $\chi_0(j) = \chi_0 \cdot (\hat{\chi} \cdot \hat{r})$, where $\chi_0$ is the user-defined input external flux in units of the Draine field \citep{Draine1978}. In order to ensure that the total field matches the desired external flux, we weight $\chi_0(j)$ by the sum of the dot products of all rays, scaling each ray with the ratio of its area to the irradiating surface angular area.

\subsection{Solving the molecular hydrogen level populations}
We have updated {\sc 3d-pdr} to solve the \ce{H2} rovibrational level populations in steady state and included \ce{H2} line cooling and collisional heating into the thermal balance.  The level density of \ce{H2} with vibrational quantum number, $v$, and rotational quantum number $J$, $n(v,J)$, is solved by the following system of rate equations in steady state:

\begin{align}
  \frac{dn(v,J)}{dt}
  &\notag =\sum_{v'J'} A(v'J' {\to} vJ)n(v',J') - n(v,J)\sum_{v'J'}A(vJ {\to} v'J') \\
  &\notag +\sum_{x\,v'J'} A(xv'J' {\to} vJ)n_x(v',J')\\
  &\notag + \sum_s\biggl(\sum_{v'J'}^{E'<E} n_s n(v'J')C_{\rm exc}(v'J' {\to} vJ) \\
  &\notag \quad\;\;+\sum_{v'J'}^{E'>E} n_s n(v'J')C_{\rm dexc}(v'J' {\to} vJ)\biggr)\\
  &\notag - n(v,J)\sum_s\biggl( \sum_{v'J'}^{E'<E} n_s C_{\rm dexc}(vJ {\to} v'J') \\
  &\notag \qquad\qquad\;+ \sum_{v'J'}^{E' > E} n_s C_{\rm exc}(vJ {\to} v'J')\biggr)\\
  &\notag -  n(v,J)\sum_{x\,v'J'} B(vJ{\to} xv'J')F_{\rm FUV}(\Delta E_{vJ,xv'J'})  \\ 
  &+ R_{\rm gr} f_{\rm gr}(v,J) - \zeta(\ce{H2})n(v,J) - \sum_m n_m k^*(\ce{H2 + m})= 0,
\end{align}

where $x\in\{B,C^+,C^-\}$ denotes the electronic levels of excited atoms and $n_x(v',J')$ their corresponding number density, while the absence of $x$ corresponds to the electronic ground state. The system includes the radiative cascade, through the Einstein $A$-coefficients, collisional de-excitation, $C_{\rm dexc}$, collisional excitation, $C_{\rm exc}$, radiative excitations due to an FUV flux, $F_{\rm FUV}$ -- here approximated using the Einstein $B$-coefficients -- excitation due to the exothermic formation on a dust grain, net destruction by cosmic-ray ionization, and destruction through two body chemical reactions between excited \ce{H2} and species, $m$, with rate $k^*(\ce{H2 + m})$. For collisions, we sum over the species $s\in\{\ce{H+}, \ce{H}, \ce{He}, \ce{o-H2}, \ce{p-H2}\}$. We assume that the exothermic formation of \ce{H2} on dust populates the levels in LTE according to the exothermic energy difference (4.4781 eV, or 51970 K) as an approximation. For collisions with \ce{H+}, we add the density of \ce{H3+}, which can undergo similar proton exchange reactions. We also include ortho-to-para conversion on dust grains \citep{LeBourlot2000}. For the model present here, we included all ground level transitions up to a maximum vibrational number, $v_{\rm max} = 10$, and maximum rotational number, $J_{\rm max} = 11$, resulting in total 132 levels. 

The effective FUV flux is scaled for ortho- and para-\ce{H2} to reproduce preferential dissociation depending on the ortho-to-para ratio. The underlying model uses the analytic results of \citet{Sternberg1999}, under the assumption that \ce{H2} mainly dissociates and fluoresces from the ground vibrational state. Given the total dissociation rate, $D$, we can solve for the dissociation rate of ortho- and para-\ce{H2}, $D_o$ and $\mathcal{D}_p$, respectively using the constraint equation:
\begin{equation}
    Dn(\ce{H2}) = D_o n(\ce{o-H2}) + D_p n(\ce{p-H2}).
\end{equation}
In the limit of low attenuation, $D_o \approx D_p$, while at high attenuation, analytically $D_p/D_o = \left ( N_p/N_o\right )^{-1/2}$, where $N_o$ and $N_p$ are the columns of \ce{o-H2} and \ce{p-H2} along the ray, respectively. Given the ortho-to-para ratio, $\psi = n(\ce{o-H2})/n(\ce{p-H2})$, 
\begin{equation}
    D_o = \frac{Dn(\ce{H2})}{n(\ce{p-H2})} \left [ \psi + \left ( \frac{N_p}{N_o}\right )^{-1/2}\right ]^{-1}
\end{equation}
and
\begin{equation}
    D_p = \frac{Dn(\ce{H2}) - n(\ce{o-H2})D_o}{n(\ce{p-H2})}.
\end{equation}
The effective FUV fields for ortho- and para-\ce{H2} can be written to satisfy these constraints
\begin{equation}
    \chi_o = \sum_k^{N_{\rm rays}} \chi_k \cdot \left ( \frac{D_o}{D}\right )
\end{equation}
and
\begin{equation}
    \chi_p = \sum_k^{N_{\rm rays}} \frac{\left( \chi_k n(\ce{H2}) - n(\ce{o-H2})\chi_o \right )}{n(\ce{p-H2})}.
\end{equation}
This scheme effectively weights the FUV field for the different components and ensures that the total excitation and dissociation rates are the same as those used in the chemistry.

The electronic levels, which provide the fluorescent excitation, are solved in steady state
\begin{equation}
\begin{split}
    \frac{dn_x(v',J')}{dt} = &\sum_{v,J} n(v,J)B(vJ\rightarrow xv'J')F_{\rm FUV}(\Delta E_{vJ,xv'J'})\\
    & - n_x(v',J')\sum_{vJ} A(xv'J' \rightarrow vJ) \\
    & - n_x(v',J')A(xv'J' \rightarrow {\rm cont}) = 0.
\end{split}
\end{equation}
We do not include transitions coupling the electronic levels, which allows the electronic rovibrational levels to be solved at the start of each iteration solely from the current ground rovibrational levels. The term $A(xv'J' \rightarrow {\rm cont})$ denotes radiative transitions to the continuum, leading to dissociation.

We exploit the fact that collisional excitation is slower than the radiative cascade and collisional de-excitation to solve the equations in steady state through a Gauss-Seidel method. We order the levels from lowest energy to highest energy. Then, starting from the highest level, we solve the level density in steady state directly, holding all other level densities constant, then iterate down the energy levels. This process is iterated until convergence. For the first call to the level population solver, the initial level populations assume LTE. For subsequent solver calls, after the chemistry solution, the initial level populations are the previous solution. The iterations are continued until a specified relative tolerance for both the level densities and the ortho-to-para ratio, of 10$^{-3}$ and 10$^{-2}$, respectively. These tolerances were tested using 1D slab models, and were found to be a good balance between computational expense and accuracy.

The \ce{H2} energy levels and line transition formatted data files are from the {\sc Cloudy} code\footnote{Use of the data has been graciously allowed with express permission via email (Ferland priv. comm.).} \citep{Shaw2005}, consisting of compiled transition line and energy level data \citep{Wolniewicz1998, Abgrall2000, Komasa2011}. We use the collisional data from \citet{Zhang2021}, which presented thermally-averaged collisional rates for a complete set of ground electronic state rovibrational energy levels.

We estimate the \ce{H2} rovibrational line emissivities assuming they are optically thin,
\begin{equation}
    \varepsilon_{ul} = \left ( \frac{1}{4\pi} \right ) n(\ce{H2})_u A_{ul} \Delta E_{ul},
\end{equation}
where $n(\ce{H2})_u$ is the density of the \ce{H2} gas in upper level $(v_u, J_u)$, $A_{ul}$ is the Einstein A-coefficient for the line, and $\Delta E_{ul} = E_u - E_l$ is the energy difference between the levels. We estimated line intensities by integrating along the line of sight, e.g., along the z-axis, accounting for local and foreground dust extinction:
\begin{equation}
    F_{ul} = \int_0^L \varepsilon_{ul} e^{-(\sigma_d N_H(<z)  - \tau_\mathrm{fg})} dz,
\end{equation}
where $\sigma_d$ is the approximate dust extinction cross section in the near infrared, $\tau_\mathrm{fg}$ is foreground dust absorption, and 
\begin{equation}
    N_H(<z) = \int_0^z n(H) dz',
\end{equation}
is the cumulative hydrogen-nuclei column density to account for local attenuation along the line-of-sight within the cloud. We use the dust extinction cross sections for $R_V = 5.5$ from \citet{Weingartner2001a, Draine2003}. For our predictions in the main text, we use $\tau_\mathrm{fg} = 2$, like that found within the Orion Bar. We note that the value for the foreground extinction scale the flux globally.

\subsection{Approximations and assumptions in the model}
Our chemical model makes several simplifying assumptions to enable a solution in three dimensions. The model currently only includes gas-phase chemistry and the formation of \ce{H2} on dust. However, since in this work we are primarily focused on highly irradiated PDR fronts, the lack of gas-grain chemistry will not noticeably impact our main results. Development of grain chemistry into {\sc 3d-pdr} is currently on-going for a future release of the code. In this paper, we use the grain-assisted ion recombination following the formalism of  \citet{Weingartner2001b}. In this treatment, positively charged ions can recombine not only via gas-phase dissociative or radiative recombination with free electrons, but also through charge exchange and electron transfer with interstellar dust grains. Throughout the density distribution, we assume a fixed grain size of $0.1\,{\mu}$m. \ce{H2} formation on the grains follows the \citet{Cazaux2002} formulation. We do not solve the radiation equation for the FUV spectrum, but instead solve for the total emission within the band \citep[c.f.][]{Flannery1980, Goicoechea2007}. Therefore, we must assume a single dust opacity for the entire spectral window. We thus also do not solve for \ce{H2} and \ce{CO} line shielding self-consistently and instead rely on precomputed shielding factors from \citet{Federman1979} (\ce{H2}) and \citet{vanDishoeck1988} (CO). While we include excited-state chemistry, we currently neglect state-dependent photoionization \citep{Ford1975}, collisional ionization, and charge transfer reactions \citep[e.g.,][for \ce{H2^* + H+ -> H2+ + H}]{Goicoechea2025b} of vibrationally excited \ce{H2}. These processes can become important in the atomic PDR and in the abundance of \ce{H2+} and \ce{H3+} in regions with a substantial population of \ce{H2^*} \citep{Goicoechea2025b}. Similar to other standard PDR codes, we assume steady-state chemistry. In {\sc 3d-pdr}, we solve the chemical network as a function of time for long timescales until the chemistry achieves a steady-state. 

While these assumptions reduce the overall self-consistency of the photochemistry, they enable a solution of the chemistry and radiation transfer in three-dimensions. Higher spatial dimensions are required to fully capture the radiation transfer of structured gas distributions, allowing the resolution of shadowing and radiation leakage. As high-performance computers become more powerful, {\sc 3d-pdr} has (and will continue to) expand the underlying chemical and radiation transfer schemes.

\subsection{Modeling an Orion Bar-like PDR}
We construct a dense PDR density distribution using the ``cavitated'' fractal cloud described above to represent a portion of the Orion Bar, with a uniform $128^3$ Cartesian grid. The box size is 0.1 pc, resulting in a resolution of 160 AU, sufficient to resolve the relevant photochemical transitions. Due to the computational limits of the three-dimensional models, we only include up to $v_{\rm max} = 10$ and $J_{\rm max} = 11$ ground electronic levels of \ce{H2} for the population solver. However, tests with one-dimensional slab models showed this is sufficient for the results presented here. 

We use an external FUV field, $\chi_0 = 3\times10^4$ in units of the Draine field \citep{Draine1978} and a constant cosmic-ray ionization rate of $\zeta = 5\times10^{-17}$ s$^{-1}$ \citep{Joblin2018}. Our results are not expected to be sensitive to deviations of $\zeta$ less than an order of magnitude. The preceding cavity is filled with a constant-density gas with hydrogen-nuclei number density $n_H \approx n(\ce{H}) + 2n(\ce{H2}) = 10^4$ cm$^{-3}$. The density structures at the front range from $10^4 - 5\times 10^6$ cm$^{-3}$. For the dust properties, we use an $R_V = 5.5$, which leads to a conversion factor between the $A_V$ and hydrogen-nuclei column density, $N(H)$ of 
$A_{V,o}=3.5\times10^{-22}\,{\rm mag}\,{\rm cm}^2$,
and an extinction coefficient in the FUV for dust absorption of $k_{\rm FUV} = 1.722$ \citep{Andree-Labsch2017, Peeters2024, Goicoechea2025}. We use a higher ray resolution level, resulting in $N_{\rm rays} = 48$ to better capture the impact of the clumps and heterogeneous structure on the radiation field. The abundances are initialized as mixed molecular hydrogen, $x(\ce{H2}) = 0.3$, $x(H) = 0.4$, with the rest of the species being atomic, $x(\ce{He}) = 0.1$, $x(\ce{C+}) = 1.4\times10^{-4}$, $x(O) = 2.6\times 10^{-4}$, $x(N) = 7.5\times 10^{-5}$ and $x(\ce{S+}) = 1.86\times10^{-5}$ \citep{Meyer1997, Sofia2004, Fuente2024, Goicoechea2025}. Electrons are initialized by charge balance.

\begin{figure*}[tb!]
    \centering
    \includegraphics[width=\textwidth]{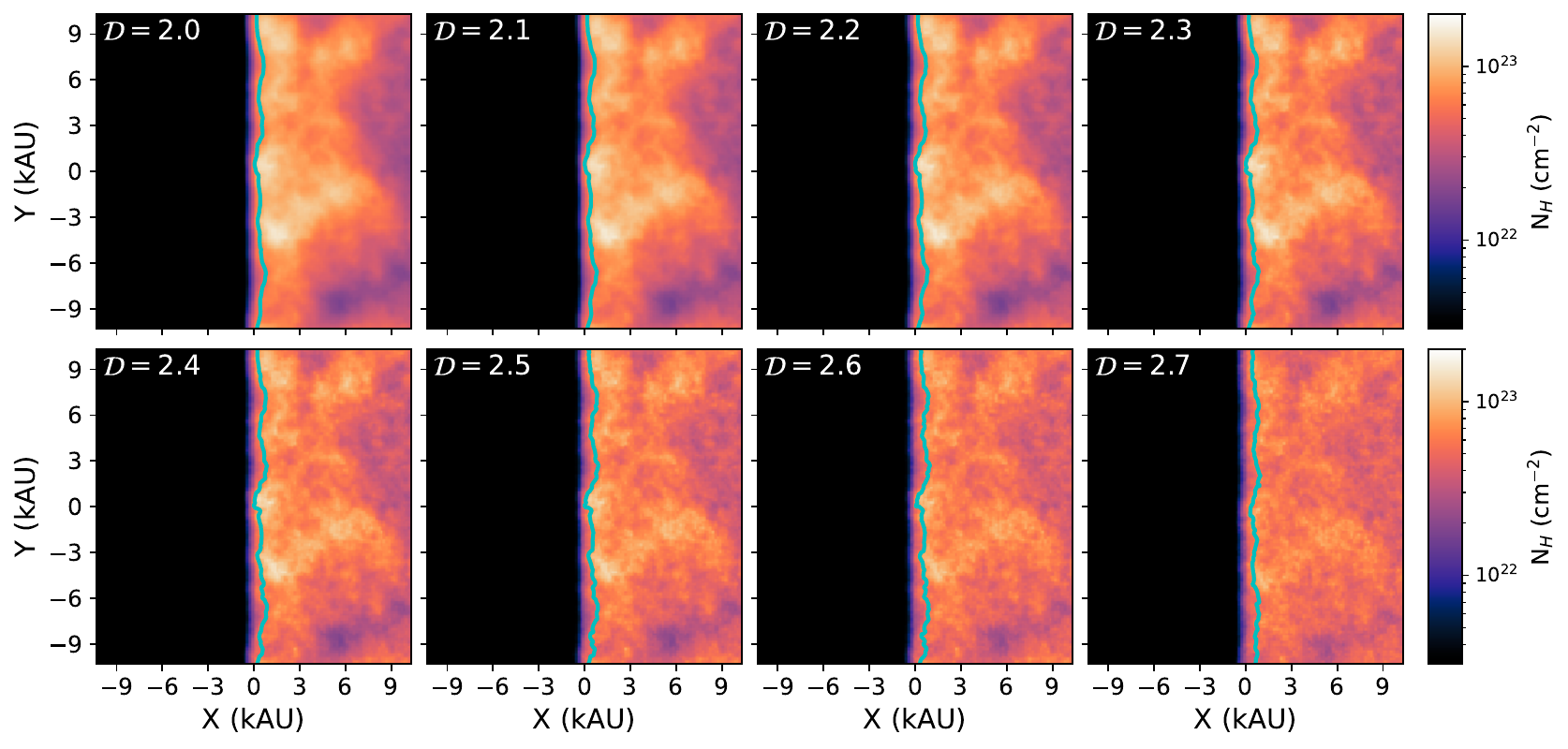}
    \caption{\label{fig:col}Hydrogen-nuclei column densities, $N_H$ (cm$^{-2}$), for the different fractal dimension distributions, denoted by the $\mathcal{D}$. The cyan line shows the line-of-sight dissociation front, $N(\ce{H2})/N_H = 0.25$.}
\end{figure*}

\section{Results}\label{sec:res}
We calculate the temperature structure and chemistry for 8 different density realizations varying the fractal dimension, $\mathcal{D}$, between 2.0 -- 2.7, to encompass a wide range of distributions from large correlated structures to a more homogeneous distribution of small clumps. Figure \ref{fig:col} shows the hydrogen-nuclei column densities for the different fractal dimensions. As the fractal dimension increases, the size of the dense substructures decreases, with low fractal dimension having high column density and large correlated structures with distinct low-density pockets. These disappear with increasing fractal dimension as the distribution becomes more homogeneous and translucent, dominated by many small clumps. 

\subsection{Thermal structure}
\begin{figure*}[htb!]
    \centering
    \includegraphics[width=\textwidth]{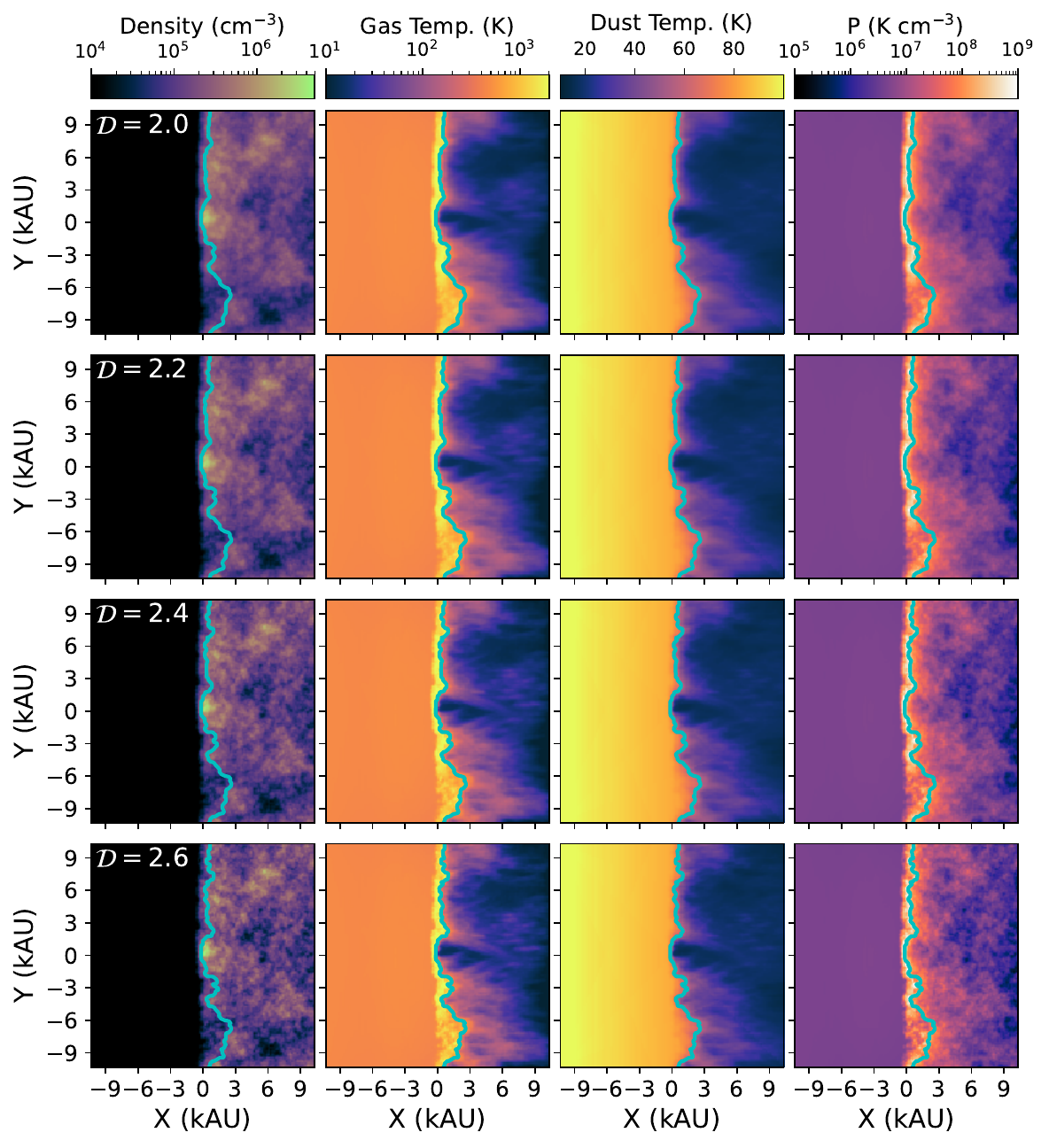}
    \caption{\label{fig:therm}Left to right: density (cm$^{-3}$), gas temperature (K), dust temperature (K), and pressure (K cm$^{-3}$) for slices through the eight fractal density distributions from top to bottom from $\mathcal{D} = 2, 2.2, 2.4$ and $2.6$. The cyan line shows the line-of-sight dissociation front, $N(\ce{H2})/N_H = 0.25$.}
\end{figure*}

Figure \ref{fig:therm} shows the gas and dust temperatures and thermal pressures for a slice through the domain for a subset of the density distributions. Preceding the PDR front, the gas temperature is a few hundred of Kelvin and the dust temperature exceeds 50~K. Exactly at the front there is an enhancement of the gas temperature to over 1000~K while the dust temperature is about 80~K. This corresponds to the enhanced pressure immediately along the dissociation front. The gas and dust temperatures rapidly drop in the cloud to tens of Kelvin for the gas and 10 Kelvin for the dust. The PDR surface pressure is marginally higher than inferred by observations ($P\approx1.38\times10^{-8}\mathrm{erg/cm^3}$ corresponding to $P/k_\text{B} = 10^{8}$ K cm$^{-3}$), but this excess is only found in a very thin band which is difficult to resolve with observations. Shadows are visible where gas and dust temperatures drop to 20~K or cooler. Below, we present the main results regarding the photochemical fronts and the impact of higher dimensionality on the FUV radiation field. 

\subsection{Molecular hydrogen ortho-to-para ratio}
\begin{figure*}[htb!]
    \centering
    \includegraphics[width=\textwidth]{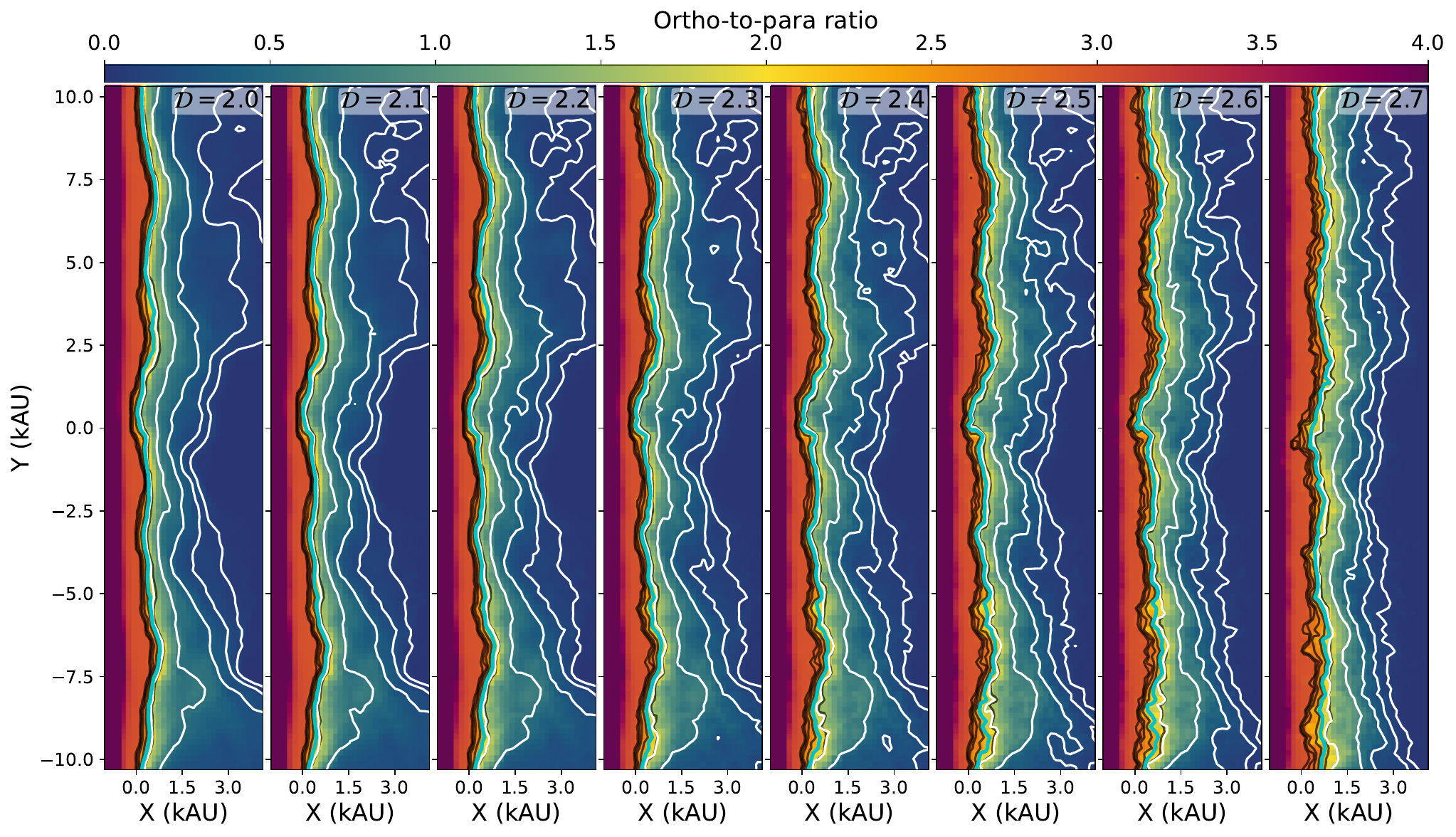}
    \caption{\label{fig:OPR}\ce{H2} Ortho-to-para ratio for the different fractal dimensions. Black and white contours denote the \ce{H2} and \ce{C+} abundance profiles, respectively. The cyan line shows the line-of-sight average \ce{H2} dissociation front.}
\end{figure*}

Figure \ref{fig:OPR} shows line-of-sight average ortho-to-para ratios (OPR) along the dissociation front. The figure shows, as contours, the abundances of \ce{H2} and \ce{C+}, and the location of the dissociation front. We find that around the \ce{H2} emission peak, the OPR is close to 3, in agreement with previous modeling of the JWST \ce{H2} lines. The OPR ratio begins to exceed 3 within the atomic gas where the \ce{H2} abundance plummets. The enhanced OPR is due to non-LTE excitation from FUV irradiation, and has been predicted previously in theory \citep{Draine1996, Sternberg1999}. An OPR exceeding 3 has not been directly observed in local regions, although it has been found in recent observations of the very-low-metallicity dwarf galaxy I Zw 18 \citep{Hunt2025}. In our models, the regions with OPR $> 3$ would not be readily observable, since they exist in the primarily atomic gas. It is also worth noting that our assumption of using a single FUV band with prescribed shielding rates may break down in these regions. Beyond the dissociation front, the OPR rapidly plummets to below 1.

\subsection{Overview of the chemical structure}
\begin{figure*}[htb!]
    \centering
    \includegraphics[width=\textwidth]{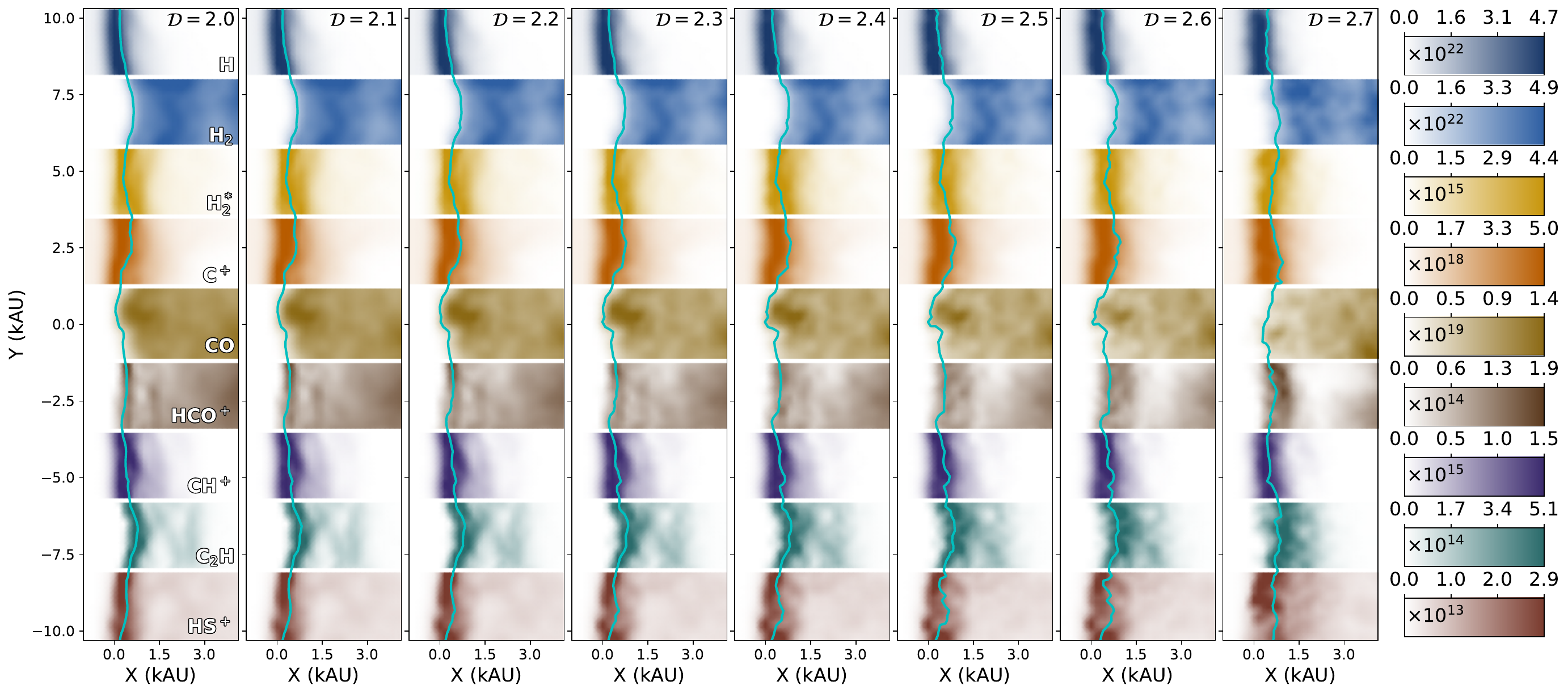}
    \caption{\label{fig:mols} Overview of the column densities of important species for the different fractal dimensions. Each specific band represents a species, annotated on the leftmost subfigure. The far right column shows the colorbars for the respective species, with the overall magnitude scale annotated inside the colorbar. The cyan line shows the line-of-sight dissociation front, $N(\ce{H2})/N_H = 0.25$.}
\end{figure*}

\begin{figure*}[htb!]
    \includegraphics[width=\textwidth]{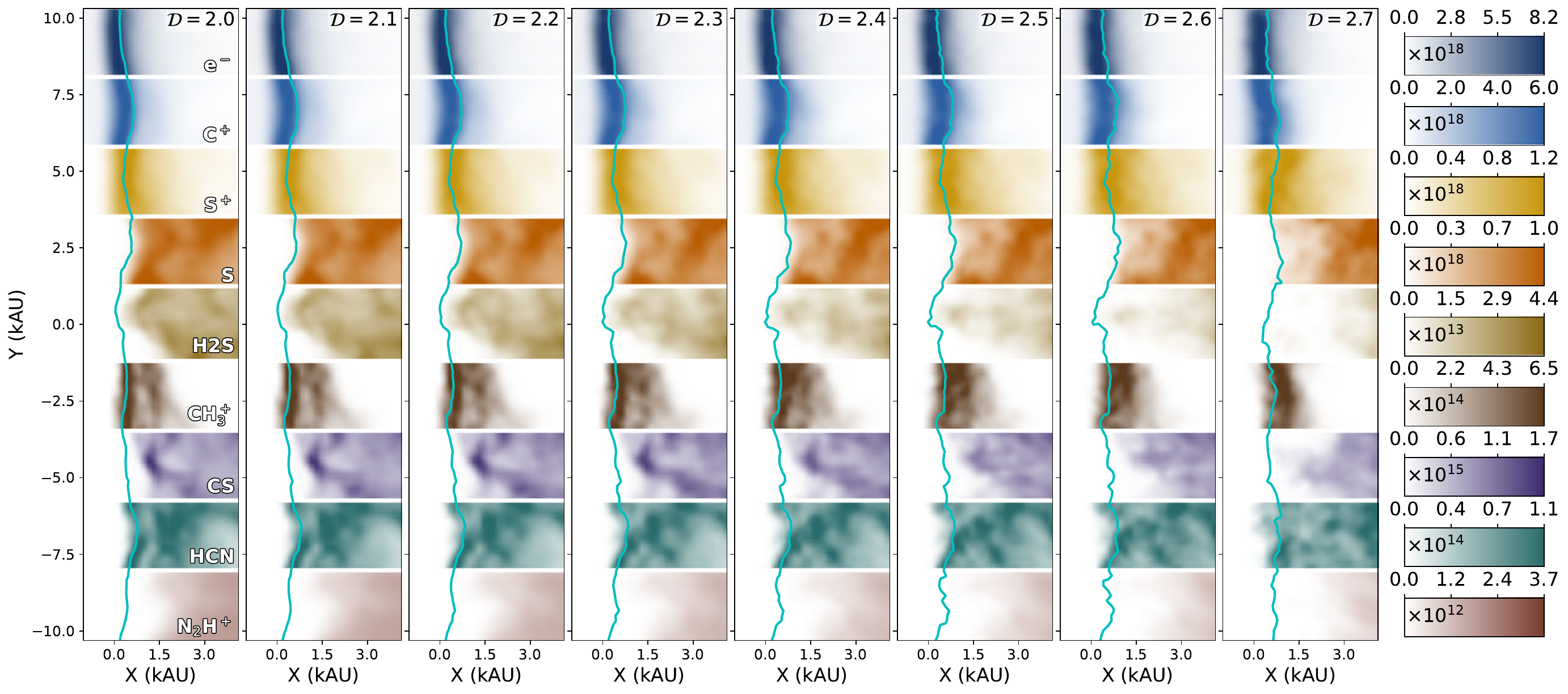}
    \caption{\label{fig:mols2}Same as Figure \ref{fig:mols} with a different set of molecules.}
\end{figure*}

Figures \ref{fig:mols} and \ref{fig:mols2} show the column densities of a number of important species and highlights the change in chemical structure of the PDR with fractal dimension. While not the focus of the present work, we present the line-of-sight average chemistry structure and defer detailed investigations into the various chemistries for latter work. 

\ce{H}, \ce{H2} and \ce{H2^*} do not morphologically change with fractal dimension. The main difference is that from larger fractal dimensions, the distribution of \ce{H2^*} slightly broadens as the gas becomes dominated by more smaller substructures, leading to more FUV excitation immediately at the front. Similarly, \ce{C+} starts tracing a broader area in front of the \ce{H2} dissociation front with higher fractal dimensions. Conversely, CO is more concentrated behind the \ce{H2} dissociation front for low fractal dimension distributions and it is more reduced for high dimension distributions. This is due to the larger more dense substructures that are dominant in lower fractal dimension distributions. 

There have been recent advances in the understanding of small hydrocarbons in strong PDRs, driven in large part by new detections in the infrared and a better understanding of excited-state chemistry \citep{Agundez2010, Zanchet2013, Berne2023, Goicoechea2025, Zannese2025b}. The column density panels highlight the role of FUV photochemistry on the chemistry of the small hydrocarbons and sulfur-bearing hydrides. \ce{CH+}, \ce{CH3+} and \ce{C2H} are tightly confined around the dissociation front, tracing its overall structure very well \citep{Nagy2013, Cuadrado2015, Nagy2015, Goicoechea2019, Zannese2025b}. At low fractal dimensions, \ce{C2H} is found in a tight band around the dissociation front, with the distribution broadening behind the front for higher fractal dimensions as the size of clumps decreases. The inverse is seen with \ce{CH+} and \ce{CH3+}, with broad features seen in the column densities due to the different large-scale structures that make multiple differentiated dissociation fronts. This seems to be in qualitative agreement with the findings of \citet{Nagy2015} that \ce{C2H} is tracing the illuminated edges of dense clumps. \ce{HCO+} has an enhancement at the dissociation front from the formation of\ce{HCO+} through vibrationally-excited hydrogen. This is in contrast to the typical formation pathway in dense, shielded clouds through \ce{CO + H3+ -> HCO+ + H2}. Low fractal dimension models, which have prominent shadows, have a substantial amount of \ce{HCO+} just behind the front. In our models, there are two main entry points for excited-state and warm hydrocarbon chemistry:
\begin{equation}
    \ce{C+ + H_2(v, J) -> CH+ + H},
\end{equation}
and 
\begin{equation}
    \ce{C2 + H2(v, J) -> C2H + H}.
\end{equation}
These reactions are highly endothermic, with endothermicities of $\Delta E/k = 4300$~K \citep{Zanchet2013} and $\Delta E/k = 1300$~K \citep{pitts1982}. At the dissociation front, the gas becomes warm enough to even start to activate the latter reaction without FUV-pumped vibrationally-excited \ce{H2}. However, it is clear that even with \ce{H2} in the first vibrational level ($v = 1$), these reactions start to become important. The combination of the warmer gas and FUV excitation pumping \ce{H2} stimulates a fast and complex hydrocarbon chemistry. Hydrocarbon ions can rapidly hydrogenate via interactions with \ce{H2}, e.g., \ce{CH+ -> CH_2+ -> CH_3+}, which undergo electron recombination to produce neutral hydrocarbons. Hydrocarbons interact with \ce{C+} to build up complexity, e.g., \ce{CH+ + C+ -> C2+ + H}.

The sulfur-bearing hydrides are also tightly associated with the dissociation front. Sulfur remains ionized deeper into the front than \ce{C+} due to the lower ionization energy. \ce{HS+} traces the immediate region around the dissociation front similar to \ce{CH+}, which broadens towards high fractal dimension. The neutral species \ce{H2S} and \ce{CS} form deeper into the molecular cloud. As the fractal dimension increases, the depth where the neutral species form also increases since the radiation encounters more homogeneously distributed clumps at the PDR front. The sulfur hydride chemistry is initiated primarily via the excited-state reaction,
\begin{equation}
    \ce{S+ + H2(v, J) -> SH+ + H, }
\end{equation}
which has an endothermicity of $\Delta E/k = 9860$~K \citep{Zanchet2019, Goicoechea2021}. Similarly to \ce{CH+}, we also see \ce{HS+} preceeding the location of the line-of-sight average \ce{H2} dissociation front, in particular for higher fractal dimensions.

\begin{figure*}[htb!]
    \centering
    \includegraphics[width=\textwidth]{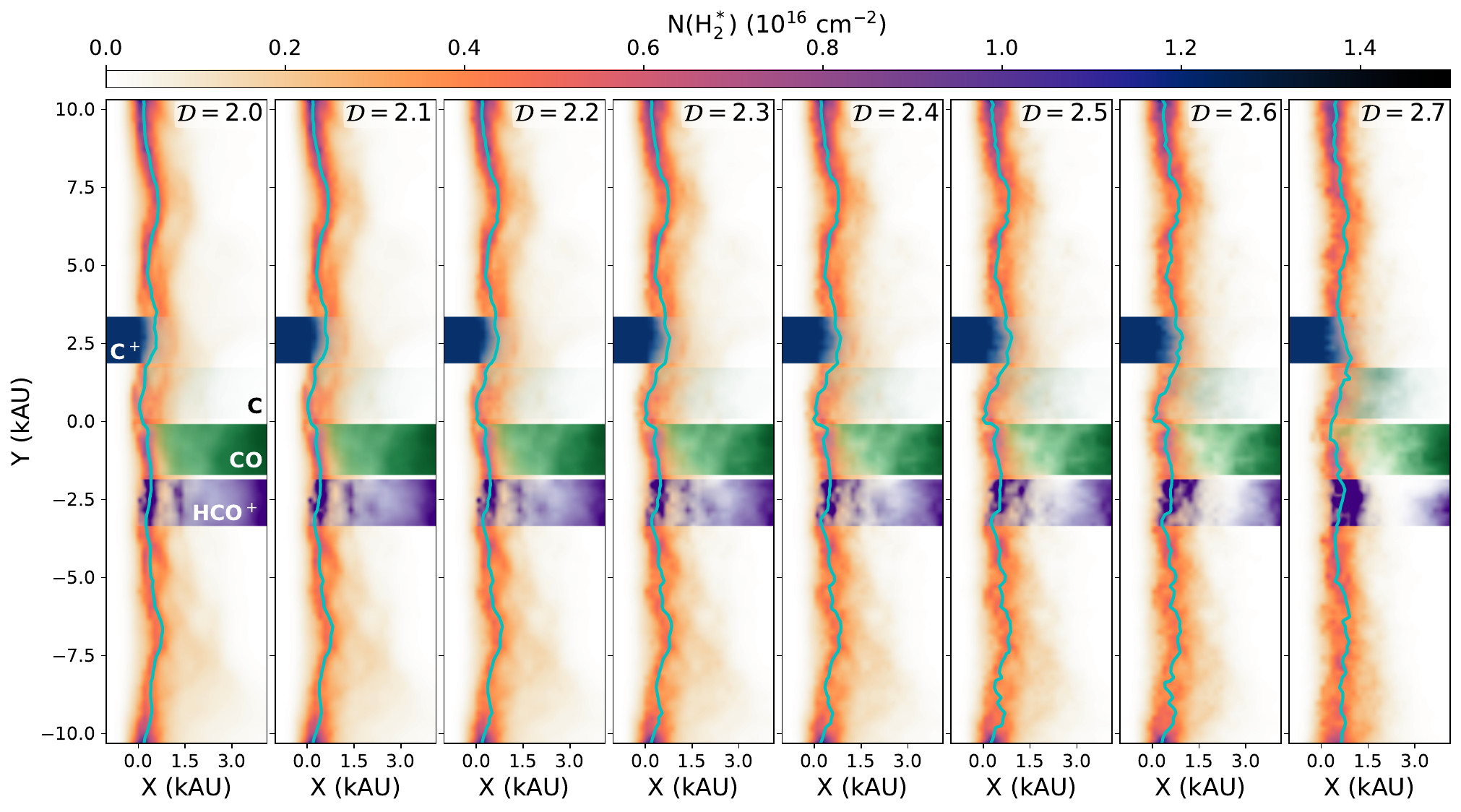}
    \caption{\label{fig:H2mColZoom} Column density of vibrationally-excited \ce{H2^*} zoomed in on the H/\ce{H2} dissociation front.  The cyan line shows the line-of-sight dissociation front, $N(\ce{H2})/N_H = 0.25$. The fractal dimension is annotated in the top right corner. The line-of-sight average abundance distributions of \ce{C+}, \ce{C}, \ce{CO}, and \ce{HCO+} are shown as the blue (top), teal, green, and purple (bottom), respectively. \ce{C+}, \ce{C} and \ce{CO} are normalized to $x_C = 1.4\times10^{-4}$, while \ce{HCO+} is normalized to $1.4\times10^{-9}$ for visual clarity.}
\end{figure*}

Figure \ref{fig:H2mColZoom} shows the average line-of-sight abundances of \ce{C+}, \ce{C}, \ce{CO} and \ce{HCO+} in strips across the dissociation front, visualized by the column density of \ce{H2^*}. This visualization helps confirm the role of the photochemistry. The \ce{C+} recombines at a location coincident to the average \ce{H2} dissociation front, with recombination beginning just before or the front. This agrees with the observations of C radio recombination preceding the \ce{H2} dissociation front \citep{Wyrowski1997, Cuadrado2019}. As seen in the figure, the recombination occurs where there is a peak in \ce{H2^*}. Therefore, there will be an ample amount of \ce{H2^*} available for reactions with \ce{C+} or other species such as \ce{C2}. For less clumpy models, there is very little neutral atomic C. The \ce{C+} either immediately forms into CO, or it is converted through excited-state chemistry into hydrocarbons. After the front, the chemistry becomes more typical of molecular cloud environments. For instance, \ce{HCO+} formation is  dominated by \ce{H3+ + CO -> HCO+ + H2}. 

\subsection{Structure of the photochemical fronts}
\begin{figure}
    \centering
    \includegraphics[width=\columnwidth]{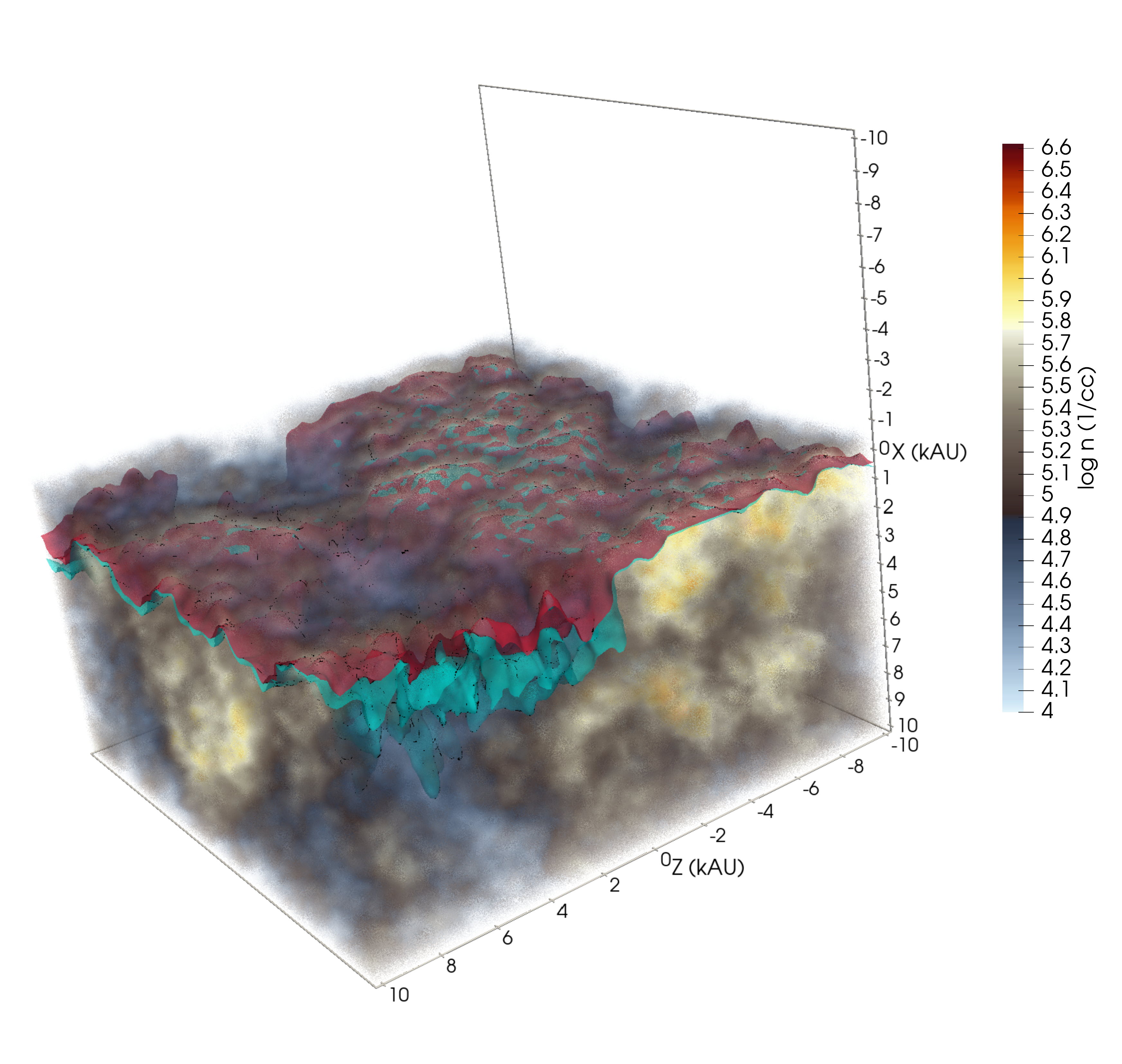}
    \caption{\label{fig:volrender} Left: Volume render of density of the $\mathcal{D} = 2.2$ distribution with the \ce{H2} dissociation front (red) and the \ce{C+} recombination front (cyan) annotated. 
    }
\end{figure}

Figure \ref{fig:volrender} shows a volume rendering of the $\mathcal{D} = 2.2$ density distribution with isocontours highlighting the location of the \ce{H2} dissociation front and the \ce{C+} recombination front. The dissociation front is visualized where $x(H) = 2x(\ce{H2}) = 0.5$ and location of the \ce{C+} recombination is where $x(\ce{C+})/x_C = 1/e$, where $x_C$ is the total elemental carbon abundance. We note that the density volume render is deliberately colored to highlight the dense substructures in the PDR. Previous PDR models of the Orion Bar have primarily utilized one-dimensional slabs, either assuming constant pressure or density, and so cannot capture the complex three-dimensional physical and chemical structure of the dissociation front shown here.

The surface of the dissociation front is highly structured with multiple incursions into the cloud as a result of low-density voids in the density distribution. There is no singular dissociation front: instead, it is a highly complex structured isosurface. Even along a perfectly edge-on line of sight, the dissociation front will not trace a planar region and instead appears structured and filamentary as the isosurface is integrated through. 

The \ce{C+} recombination front (traced by the cyan isocontour in Figure \ref{fig:volrender}) aligns closely with the \ce{H2} dissociation front, confirming their co-spatiality. However, while they are broadly coincident, they do not completely overlap; in certain regions, the cyan isocontours extend beyond the red \ce{H2} boundary, indicating localized offsets in the transition zones. Voids in the density distribution enable significant incursions of \ce{C+} spatially, filling their volume. The result are regions where \ce{C+} and \ce{H2} fully co-exist, driving excited-state hydrocarbon chemistry, as demonstrated above.

We can also examine the structure of the \ce{H2} dissociation front by looking at the distribution of the FUV-pumped vibrationally-excited molecular hydrogen, \ce{H2^*}. The \ce{H2^*} column density, shown in Figure \ref{fig:H2mColZoom}, exhibits a complex structure and is not entirely confined within a tight, narrow region. Furthermore, it extends deeper into the cloud in regions where there are voids, exceeding 2 kAU away in some distributions. Interestingly, a void along the line-of-sight in the $\mathcal{D} = 2.0$ and $2.1$ distributions produces arc-like and filamentary features in projection tracing its surface. In all cases, the average \ce{H2} dissociation front is relatively aligned with the peak of the \ce{H2^*} column density, and the \ce{H2^*} peak is confined to a region less than 2 kAU in extent. At the distance of the Orion Bar, this corresponds to less than 5\arcsec across.

\subsection{Separation of the H- and C- fronts}
As discussed above, the \ce{H2} dissociation front is extended over a region of about 2 kAU. However, the separation between the \ce{H2} dissociation and \ce{C+} recombination front is often reported as a singular number. For each line-of-sight parallel to the irradiation direction, we take the separation of the C- and H- fronts to be the distance between  where $x(\ce{H2}) = 0.25$ and $x(\ce{C+})/x_C = 1/e$, resulting in a distribution of $128^2$ front separations. Observationally, this is typically traced with NIR \ce{H2} emission, such as the (1-0)S(1) rovibrational line at 2.12 $\mu$m, which should coincide with the the enhancement of $x(\ce{H_2^*})$. We have tested this, and found that while it marginally shifts the overall values, it does not change the results. We chose to use $x(\ce{H2}) = 0.25$ since it is the more consistent tracer for the chemistry. 

\begin{figure}
    \centering
    \includegraphics[width=\columnwidth]{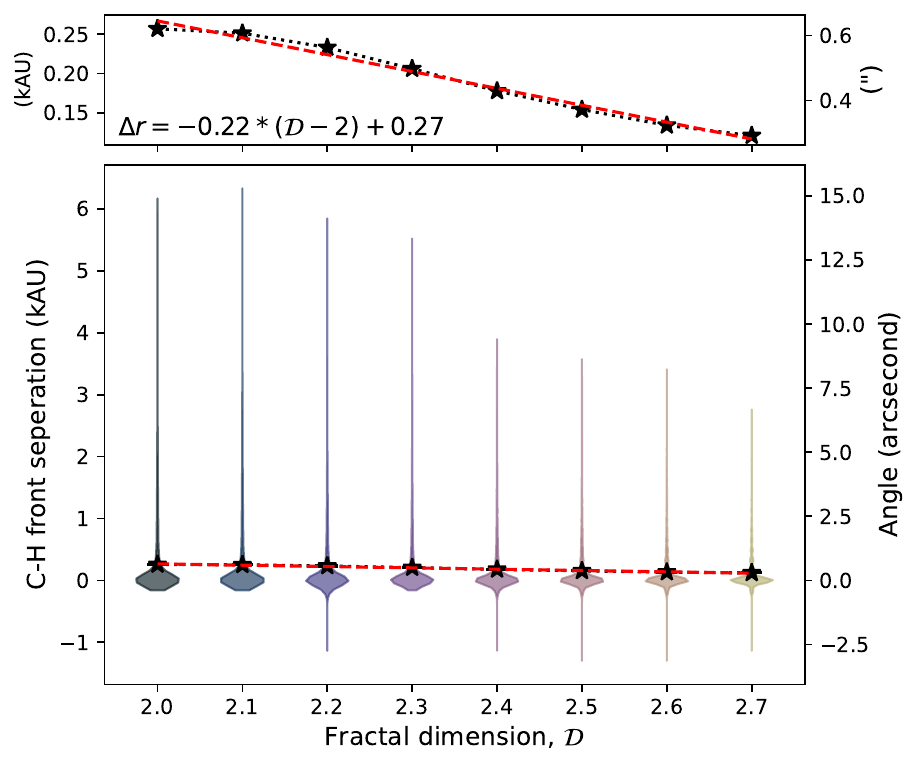}
    \caption{\label{fig:CHhist}Main: Violin plots showing the distribution of the separation of the \ce{C+}/C/CO and H/\ce{H2} dissociation fronts, in milliparsec or arcsecond (assuming the distance to the Orion Bar, 414 pc), as a function of fractal dimension. The stars denote the mean and the red line shows the fit annotated in the top panel. Top: Zooms in to highlight the trend with the means separation distances in kAU and arcsecond.}
\end{figure}

Figure \ref{fig:CHhist} shows the distribution of the C-H front distances as a function of fractal dimension via violin histogram plots. The separation distributions are centered on cospatiality, $\delta r_{C-H} \approx 0$, with a broad tail out to 6 kAU, corresponding to separations of up to 15 arcseconds for the Orion Bar. The peak of this distribution is $\delta r_{C-H} < 0.25$ kAU ($< 1$ arcsecond), with the mean consistent with cospatiality to the current observational limits. Finally, a trend with fractal dimension can be identified: as the cloud structures become more homogeneous in size and smaller, the separation distribution shrinks and becomes tighter around cospatial. 

These distributions firmly demonstrate that the fronts are effectively co-spatial. Most interestingly, we find that in higher fractal dimension distributions, there are regions where $\delta r_{C-H} < 0$, such that the C-front precedes the H front. The C-front preceding or cospatial with the H-front was hinted by observations of \ce{C} radio recombination lines in \citep{Wyrowski1997, Cuadrado2019}. Preceeding and/or cospatial fronts have been predicted also by high-temperature/high-pressure one-dimensional PDR models \citep[][]{Joblin2018, Goicoechea2025, Goicoechea2025b}. By including heterogeneous gas distributions in three dimensions, we have demonstrated that higher-density substructures can naturally produce C-fronts that are cospatial with or precede the \ce{H2} front without invoking complex chemodynamical processes.

\subsection{Near-infrared molecular hydrogen emission at the dissociation front}
We have computed synthetic flux emission maps for the commonly used (1-0)S(1) \ce{H2} rovibrational line at 2.12 $\mu$m and (0-0)S(9) at 4.69 $\mu$m. These lines correspond to the F212N and F470N filters equipped on the JWST NIRCAM imager, respectively. We assumed the line emission is optically thin and accounted for internal dust attenuation within the domain along the line of sight. Furthermore, we applied a foreground extinction correction of $\tau_\mathrm{fg} = 2$ to account for intervening gas, consistent with recent JWST PDRs4All observations of the Orion Bar \citep{Peeters2024}. The foreground extinction only changes the overall strength of the ``observed'' emission. Since we solve for the level populations of \ce{H2} directly, the line emissivities are self-consistently calculated.

\begin{figure*}[htb!]
    \centering
    \includegraphics[width=\textwidth]{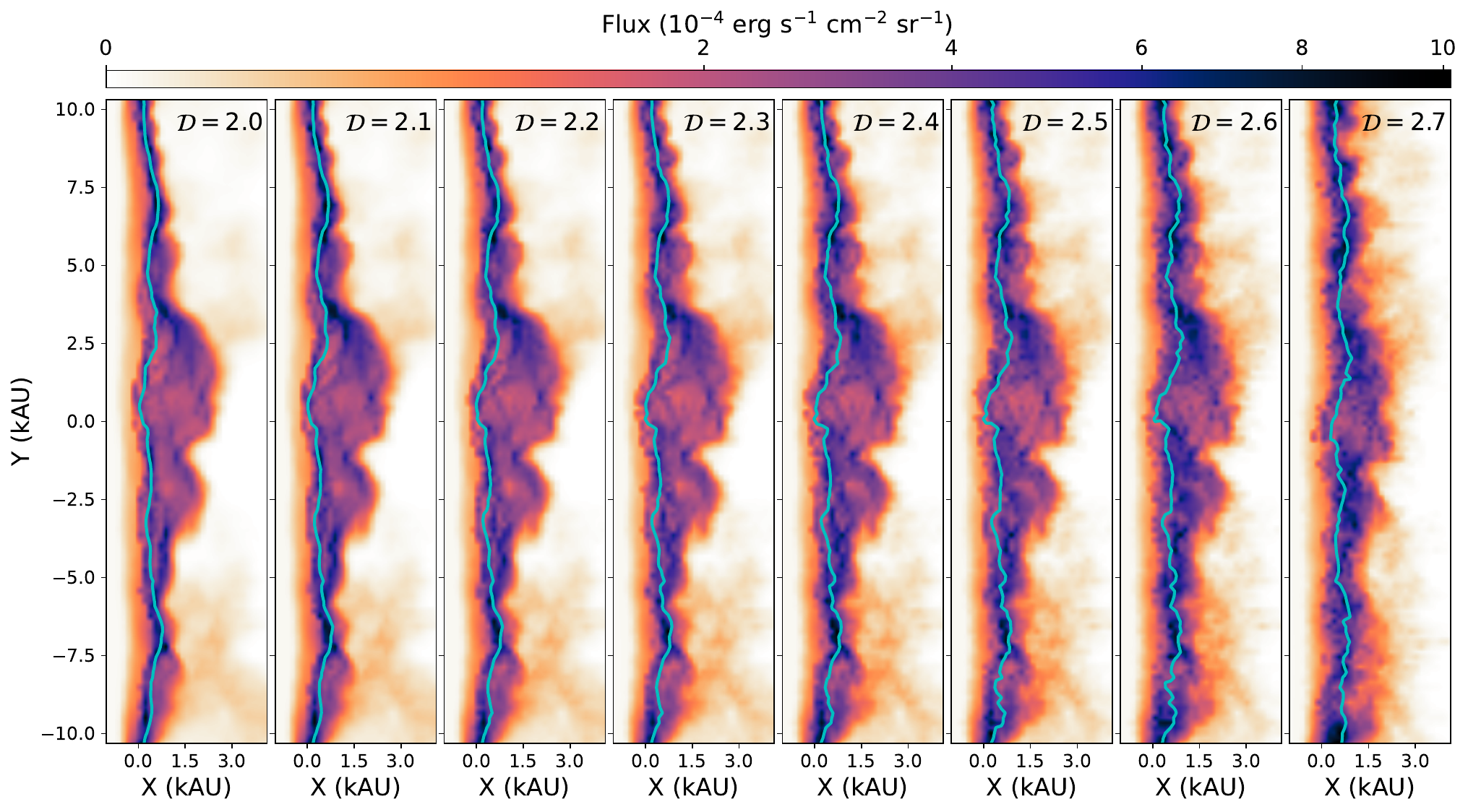}
    \caption{\label{fig:H2_O1} Line intensities of the (1-0)S(1) line at 2.12 $\mu$m for the different fractal dimensions, zoomed in at the PDR interface. The cyan line shows the line-of-sight dissociation front, $N(\ce{H2})/N_H = 0.25$. Color map has been stretched using a square-root scaling.}
\end{figure*}

\begin{figure*}[htb!]
    \centering
    \includegraphics[width=\textwidth]{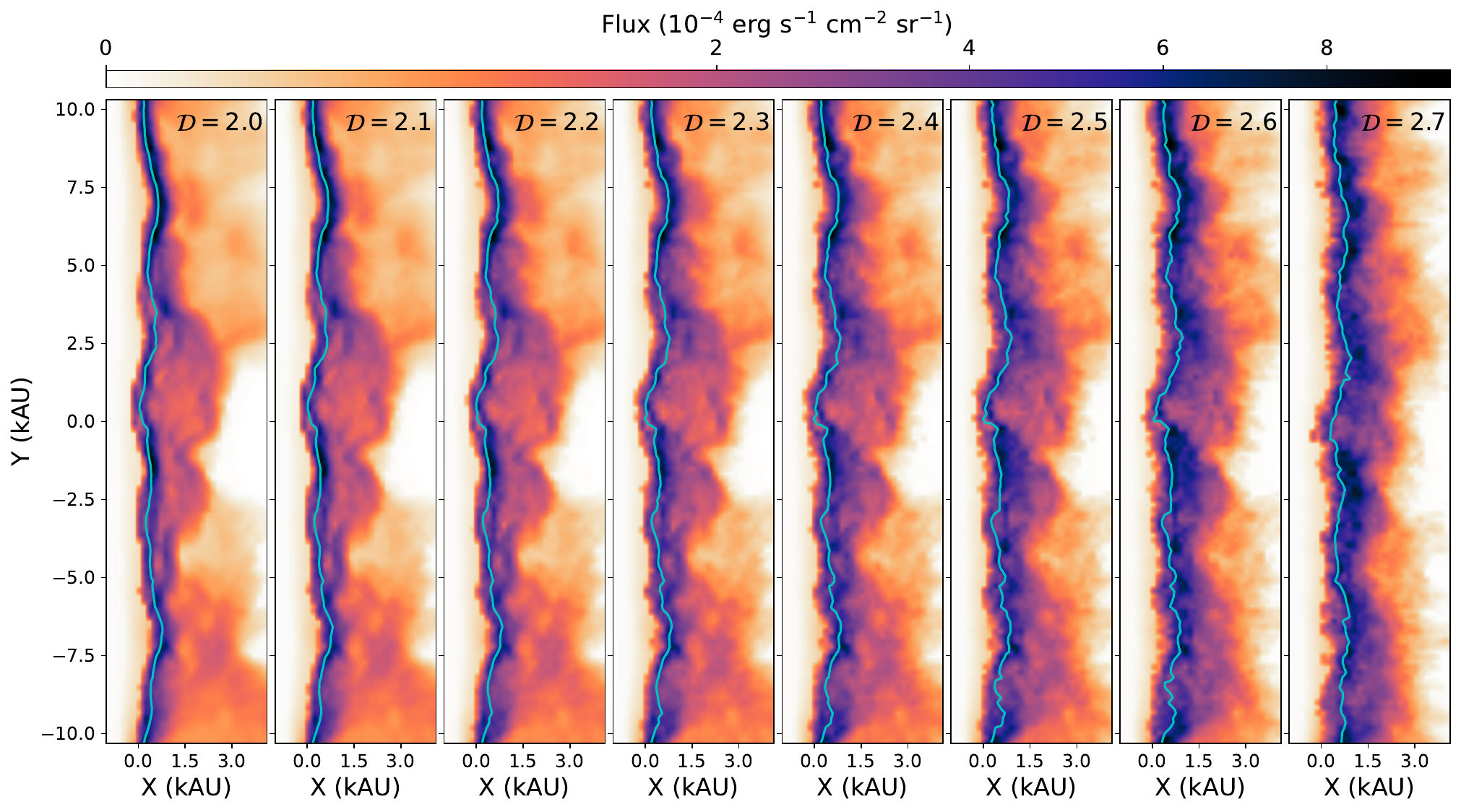}
    \caption{\label{fig:H2_s9} Same as Figure \ref{fig:H2_O1} but for the (0-0)S(9) line at 4.69 $\mu$m.}
\end{figure*}

Figure \ref{fig:H2_O1} shows the line intensity at the PDR surface for different values of the fractal dimension. The values of the intensity we predict are similar to that observed in the Orion Bar as presented by the PDRs4All team (\citet{Habart2024} and see in particular Fig. 7 and Table 3 of \citet{Peeters2024}) and observed on the ground with Keck/NIRC2 \citep{Habart2023}. These observations show line intensities around $2-7\times 10^{-4}$ erg s$^{-1}$ cm$^{-2}$ sr$^{-1}$ in the Bar. Our lower fractal dimension distributions are consistent with this range. The figure shows that the emission line, as expected, traces the PDR front very well and where \ce{H2^*} is abundant. Due to this, we find a clear trend with varying fractal dimension. For low fractal dimension, the presence of large structured voids are visible as incursions into the cloud that appear as filaments and arcs. As the PDR surface becomes increasingly clumpy and loses coherent structure at higher fractal dimensions, these complex features disappear, causing the emission to collapse into a more linear, but clumpy, PDR front. Although weaker, we find that the high rotational line (0-0)S(9), shown in Figure \ref{fig:H2_s9}, exhibits similar features. Our predicted line intensities are close to those seen in the recent JWST observations, although a bit elevated. Our line intensities may be elevated since the model's total hydrogen-nuclei column density is slightly elevated compared to observational constraints \citep[see][]{Hogerheijde1995, Berne2014, Cuadrado2015}. These results represent new self-consistent 3D astrochemical emission maps of \ce{H2}, marking a significant advancement in the spatial modeling of such regions.

\section{Discussion and conclusions}\label{sec:disc}
We have presented a novel high-resolution three-dimensional chemical model and \ce{H2} emission maps of a dense PDR, focusing on the separation of the C- and H-fronts and the morphology of these fronts. This separation has been a topic of intense interest since the high-resolution ALMA observations of \citep{Goicoechea2016} and the unmatched spatial resolution and spectral information provided by the program PDRs4All by the James Webb Space Telescope \citep{Habart2024, Peeters2024, Goicoechea2025}. The complex multiple dissociation fronts and tiered structure seen in observations has posed a change for chemical models produced by the current 1D state-of-the-art PDR models. Resolving this problem and understanding the morphology of the Orion Bar is crucial, particularly since this object has been the prototype benchmark for PDRs for decades \citep{Roellig2007}. 

The main impact of higher spatial dimensionality is due to the propagation of FUV radiation and the escape of line radiation for cooling. In one-dimensional models, the attenuating column density is calculated solely by an integration along a pencil beam aligned with the main spatial axis. In three-dimensional plane-parallel models, the irradiation is not from a pencil beam but from one of the domain surfaces. Therefore, the radiation power is spread across the entire surface, and the attenuating column density requires taking into account all lines of sight towards the irradiating plane. At very low column densities in the uniform density preceding cavity, these results converge. However, with increasing distance, and when non-uniform density structures are present, this is no longer the case. With non-uniform density distributions, one line of sight (e.g., along the -x direction) may pass through a high-density core, while another, e.g., 30$^\circ$ away, may pass through a low-density void. These differences help produce the complex features at the dissociation fronts, with some lines of sight having significantly higher UV fields for their position, and others experiencing much dimmer UV fields. Furthermore, the inclusion of the three-dimensional structure leads to shadowing, such that high dense features at the surface of the front block the FUV flux, shielding gas behind it. Since the surface is not uniform, the shadowing is anisotropic. The net effect allows the merging of the H- and C-fronts with a complex front structure exhibited in spatially-resolved observations. While the clumpiness can be modeled through statistical clump distributions, shadowing and filamentary features cannot be readily reproduced through these means.

Observations with the JWST NIRCAM show that the morphology of the gas density at the surface is complex as a result of the interplay between the hydrodynamics, radiation field, and magnetic field. Our three-dimensional model unambiguously demonstrates that these fronts can not only be cospatial, but that the C-front can precede the \ce{H2} front. Furthermore, the predicted line emission maps show that the filamentary and arc structures seen in observed maps may be the result of the complex shape of the dissociation front observed edge on \citep{Habart2023, Habart2024}. The combination of the preceding C-fronts, the observed filamentary and arc structures in \ce{H2} emission and overall magnitude of the emission give preference to fractal models between $\mathcal{D} \approx 2.2 - 2.4$. Lower values of $\mathcal{D}$ provide very large correlated structures that do not lead to there being a statistically significant amount of C-fronts preceding H-fronts. Conversely, high values of $\mathcal{D}$ exhibit purely clumpy and very strong emission distributions. Therefore, fractal dimensions between this are preferred, with a mix of more large-scale features, such as those produced by compressive motions, with turbulent mixing.

We note that while the main objective was not to provide an optimized fit to the Orion Bar, our three-dimensional chemical model was able to reproduce many of the features found in observations of the Bar. We, thus, conclude that the spatial dimensionality of the PDR fronts is a key factor in the photochemistry at the dissociation fronts and the spatial distribution of the emission at these fronts. Future works can now investigate gas morphologies induced by different physical mechanisms and examine their photochemistry and emissions in three dimensions.

\begin{acknowledgements}
We thank Javier Goicoechea for the interesting discussions and for providing a sulfur subnetwork that significantly improved the chemical network utilized. We are greatful for the helpful comments from the anonymous referee. BALG is supported by the German Research Foundation (DFG) in the form of an Emmy Noether Research Group - DFG project \#542802847 (GA 3170/3-1). TGB acknowledges support from the Leading Innovation and Entrepreneurship Team of Zhejiang Province of China (Grant No. 2023R01008). The authors gratefully acknowledge the computing time granted by the Center for Computational Sciences and Simulation (CCSS) of the University of Duisburg-Essen and provided on the supercomputer amplitUDE (DFG project 459398823; grant ID INST 20876/423-1 FUGG ) at the Center for Information and Media Services (ZIM). 
\end{acknowledgements}

\bibliographystyle{aa}
\bibliography{lib} 

@ARTICLE{Menten2007,
       author = {{Menten}, K.~M. and {Reid}, M.~J. and {Forbrich}, J. and {Brunthaler}, A.},
        title = "{The distance to the Orion Nebula}",
      journal = {\aap},
         year = 2007,
        month = nov,
       volume = {474},
       number = {2},
        pages = {515-520},
          doi = {10.1051/0004-6361:20078247},
archivePrefix = {arXiv},
       eprint = {0709.0485},
 primaryClass = {astro-ph},
       adsurl = {https://ui.adsabs.harvard.edu/abs/2007A&A...474..515M}
}

@ARTICLE{Zhu26,
       author = {{Zhu}, Zhengping and {Bisbas}, Thomas G. and {Tang}, Xuefei and {Gaches}, Brandt A.~L. and {Zhang}, Tianwei and {Chen}, Huaxi},
        title = "{RAYTHEIA: a high-performance ray-tracing algorithm for three-dimensional direction-dependent equations in astronomical simulations}",
      journal = {\mnras},
         year = 2026,
        month = jun,
       volume = {549},
       number = {2},
          eid = {stag906},
        pages = {stag906},
          doi = {10.1093/mnras/stag906},
archivePrefix = {arXiv},
       eprint = {2605.09882},
 primaryClass = {astro-ph.IM},
       adsurl = {https://ui.adsabs.harvard.edu/abs/2026MNRAS.549ag906Z}
}

@ARTICLE{Roellig2007,
       author = {{R{\"o}llig}, M. and {Abel}, N.~P. and {Bell}, T. and {Bensch}, F. and {Black}, J. and {Ferland}, G.~J. and {Jonkheid}, B. and {Kamp}, I. and {Kaufman}, M.~J. and {Le Bourlot}, J. and {Le Petit}, F. and {Meijerink}, R. and {Morata}, O. and {Ossenkopf}, V. and {Roueff}, E. and {Shaw}, G. and {Spaans}, M. and {Sternberg}, A. and {Stutzki}, J. and {Thi}, W.-F. and {van Dishoeck}, E.~F. and {van Hoof}, P.~A.~M. and {Viti}, S. and {Wolfire}, M.~G.},
        title = "{A photon dominated region code comparison study}",
      journal = {\aap},
         year = 2007,
        month = may,
       volume = {467},
       number = {1},
        pages = {187-206},
          doi = {10.1051/0004-6361:20065918},
archivePrefix = {arXiv},
       eprint = {astro-ph/0702231},
 primaryClass = {astro-ph},
       adsurl = {https://ui.adsabs.harvard.edu/abs/2007A&A...467..187R}
}

@ARTICLE{Pomelnikov2025,
       author = {{Pomelnikov}, Ivan A. and {Riashchikov}, Dmitrii S. and {Zavershinskii}, Dmitrii I. and {Molevich}, Nonna E.},
        title = "{Isentropic instability and dynamic substructures in the Orion Bar photodissociation region: Analytical and numerical insights}",
      journal = {\aap},
         year = 2025,
        month = sep,
       volume = {701},
          eid = {A59},
        pages = {A59},
          doi = {10.1051/0004-6361/202554479},
       adsurl = {https://ui.adsabs.harvard.edu/abs/2025A&A...701A..59P}
}

@ARTICLE{Goicoechea2025,
       author = {{Goicoechea}, J.~R. and {Pety}, J. and {Cuadrado}, S. and {Bern{\'e}}, O. and {Dartois}, E. and {Gerin}, M. and {Joblin}, C. and {K{\l}os}, J. and {Lique}, F. and {Onaka}, T. and {Peeters}, E. and {Tielens}, A.~G.~G.~M. and {Alarc{\'o}n}, F. and {Bron}, E. and {Cami}, J. and {Canin}, A. and {Chapillon}, E. and {Chown}, R. and {Fuente}, A. and {Habart}, E. and {Kannavou}, O. and {Le Petit}, F. and {Santa-Maria}, M.~G. and {Schroetter}, I. and {Sidhu}, A. and {Trahin}, B. and {Van De Putte}, D. and {Zannese}, M.},
        title = "{PDRs4All: XII. Far-ultraviolet-driven formation of simple hydrocarbon radicals and their relation with polycyclic aromatic hydrocarbons}",
      journal = {\aap},
         year = 2025,
        month = apr,
       volume = {696},
          eid = {A100},
        pages = {A100},
          doi = {10.1051/0004-6361/202453350},
archivePrefix = {arXiv},
       eprint = {2503.03353},
 primaryClass = {astro-ph.GA},
       adsurl = {https://ui.adsabs.harvard.edu/abs/2025A&A...696A.100G}
}

@ARTICLE{Goicoechea2025b,
       author = {{Goicoechea}, Javier R. and {Roncero}, Octavio and {Roueff}, Evelyne and {Black}, John H. and {Schroetter}, Ilane and {Bern{\'e}}, Olivier},
        title = "{H3+ in irradiated protoplanetary disks: Linking far-ultraviolet radiation and water vapor}",
      journal = {\aap},
         year = 2025,
        month = nov,
       volume = {703},
          eid = {A189},
        pages = {A189},
          doi = {10.1051/0004-6361/202555842},
archivePrefix = {arXiv},
       eprint = {2506.05189},
 primaryClass = {astro-ph.GA},
       adsurl = {https://ui.adsabs.harvard.edu/abs/2025A&A...703A.189G}
}

@ARTICLE{Goicoechea2021,
       author = {{Goicoechea}, J.~R. and {Aguado}, A. and {Cuadrado}, S. and {Roncero}, O. and {Pety}, J. and {Bron}, E. and {Fuente}, A. and {Riquelme}, D. and {Chapillon}, E. and {Herrera}, C. and {Duran}, C.~A.},
        title = "{Bottlenecks to interstellar sulfur chemistry. Sulfur-bearing hydrides in UV-illuminated gas and grains}",
      journal = {\aap},
         year = 2021,
        month = mar,
       volume = {647},
          eid = {A10},
        pages = {A10},
          doi = {10.1051/0004-6361/202039756},
archivePrefix = {arXiv},
       eprint = {2101.01012},
 primaryClass = {astro-ph.GA},
       adsurl = {https://ui.adsabs.harvard.edu/abs/2021A&A...647A..10G}
}

@ARTICLE{Peeters2024,
       author = {{Peeters}, Els and {Habart}, Emilie and {Bern{\'e}}, Olivier and {Sidhu}, Ameek and {Chown}, Ryan and {Van De Putte}, Dries and {Trahin}, Boris and {Schroetter}, Ilane and {Canin}, Am{\'e}lie and {Alarc{\'o}n}, Felipe and {Schefter}, Bethany and {Khan}, Baria and {Pasquini}, Sofia and {Tielens}, Alexander G.~G.~M. and {Wolfire}, Mark G. and {Dartois}, Emmanuel and {Goicoechea}, Javier R. and {Maragkoudakis}, Alexandros and {Onaka}, Takashi and {Pound}, Marc W. and {Vicente}, S{\'\i}lvia and {Abergel}, Alain and {Bergin}, Edwin A. and {Bernard-Salas}, Jeronimo and {Boersma}, Christiaan and {Bron}, Emeric and {Cami}, Jan and {Cuadrado}, Sara and {Dicken}, Daniel and {Elyajouri}, Meriem and {Fuente}, Asunci{\'o}n and {Gordon}, Karl D. and {Issa}, Lina and {Joblin}, Christine and {Kannavou}, Olga and {Lacinbala}, Ozan and {Languignon}, David and {Le Gal}, Romane and {Meshaka}, Raphael and {Okada}, Yoko and {Robberto}, Massimo and {R{\"o}llig}, Markus and {Schirmer}, Thi{\'e}baut and {Tabone}, Benoit and {Zannese}, Marion and {Aleman}, Isabel and {Allamandola}, Louis and {Auchettl}, Rebecca and {Baratta}, Giuseppe Antonio and {Bejaoui}, Salma and {Bera}, Partha P. and {Black}, John H. and {Boulanger}, Francois and {Bouwman}, Jordy and {Brandl}, Bernhard and {Brechignac}, Philippe and {Br{\"u}nken}, Sandra and {Buragohain}, Mridusmita and {Burkhardt}, Andrew and {Candian}, Alessandra and {Cazaux}, St{\'e}phanie and {Cernicharo}, Jose and {Chabot}, Marin and {Chakraborty}, Shubhadip and {Champion}, Jason and {Colgan}, Sean W.~J. and {Cooke}, Ilsa R. and {Coutens}, Audrey and {Cox}, Nick L.~J. and {Demyk}, Karine and {Meyer}, Jennifer Donovan and {Foschino}, Sacha and {Garc{\'\i}a-Lario}, Pedro and {Gerin}, Maryvonne and {Gottlieb}, Carl A. and {Guillard}, Pierre and {Gusdorf}, Antoine and {Hartigan}, Patrick and {He}, Jinhua and {Herbst}, Eric and {Hornekaer}, Liv and {J{\"a}ger}, Cornelia and {Janot-Pacheco}, Eduardo and {Kaufman}, Michael and {Kendrew}, Sarah and {Kirsanova}, Maria S. and {Klaassen}, Pamela and {Kwok}, Sun and {Labiano}, {\'A}lvaro and {Lai}, Thomas S.-Y. and {Lee}, Timothy J. and {Lefloch}, Bertrand and {Le Petit}, Franck and {Li}, Aigen and {Linz}, Hendrik and {Mackie}, Cameron J. and {Madden}, Suzanne C. and {Mascetti}, Jo{\"e}lle and {McGuire}, Brett A. and {Merino}, Pablo and {Micelotta}, Elisabetta R. and {Misselt}, Karl and {Morse}, Jon A. and {Mulas}, Giacomo and {Neelamkodan}, Naslim and {Ohsawa}, Ryou and {Paladini}, Roberta and {Palumbo}, Maria Elisabetta and {Pathak}, Amit and {Pendleton}, Yvonne J. and {Petrignani}, Annemieke and {Pino}, Thomas and {Puga}, Elena and {Rangwala}, Naseem and {Rapacioli}, Mathias and {Ricca}, Alessandra and {Roman-Duval}, Julia and {Roser}, Joseph and {Roueff}, Evelyne and {Rouill{\'e}}, Ga{\"e}l and {Salama}, Farid and {Sales}, Dinalva A. and {Sandstrom}, Karin and {Sarre}, Peter and {Sciamma-O'Brien}, Ella and {Sellgren}, Kris and {Shenoy}, Sachindev S. and {Teyssier}, David and {Thomas}, Richard D. and {Togi}, Aditya and {Verstraete}, Laurent and {Witt}, Adolf N. and {Wootten}, Alwyn and {Ysard}, Nathalie and {Zettergren}, Henning and {Zhang}, Yong and {Zhang}, Ziwei E. and {Zhen}, Junfeng},
        title = "{PDRs4All: III. JWST's NIR spectroscopic view of the Orion Bar}",
      journal = {\aap},
         year = 2024,
        month = may,
       volume = {685},
          eid = {A74},
        pages = {A74},
          doi = {10.1051/0004-6361/202348244},
archivePrefix = {arXiv},
       eprint = {2310.08720},
 primaryClass = {astro-ph.GA},
       adsurl = {https://ui.adsabs.harvard.edu/abs/2024A&A...685A..74P}
}

@ARTICLE{Habart2024,
       author = {{Habart}, Emilie and {Peeters}, Els and {Bern{\'e}}, Olivier and {Trahin}, Boris and {Canin}, Am{\'e}lie and {Chown}, Ryan and {Sidhu}, Ameek and {Van De Putte}, Dries and {Alarc{\'o}n}, Felipe and {Schroetter}, Ilane and {Dartois}, Emmanuel and {Vicente}, S{\'\i}lvia and {Abergel}, Alain and {Bergin}, Edwin A. and {Bernard-Salas}, Jeronimo and {Boersma}, Christiaan and {Bron}, Emeric and {Cami}, Jan and {Cuadrado}, Sara and {Dicken}, Daniel and {Elyajouri}, Meriem and {Fuente}, Asunci{\'o}n and {Goicoechea}, Javier R. and {Gordon}, Karl D. and {Issa}, Lina and {Joblin}, Christine and {Kannavou}, Olga and {Khan}, Baria and {Lacinbala}, Ozan and {Languignon}, David and {Le Gal}, Romane and {Maragkoudakis}, Alexandros and {Meshaka}, Raphael and {Okada}, Yoko and {Onaka}, Takashi and {Pasquini}, Sofia and {Pound}, Marc W. and {Robberto}, Massimo and {R{\"o}llig}, Markus and {Schefter}, Bethany and {Schirmer}, Thi{\'e}baut and {Tabone}, Benoit and {Tielens}, Alexander G.~G.~M. and {Wolfire}, Mark G. and {Zannese}, Marion and {Ysard}, Nathalie and {Miville-Deschenes}, Marc-Antoine and {Aleman}, Isabel and {Allamandola}, Louis and {Auchettl}, Rebecca and {Baratta}, Giuseppe Antonio and {Bejaoui}, Salma and {Bera}, Partha P. and {Black}, John H. and {Boulanger}, Francois and {Bouwman}, Jordy and {Brandl}, Bernhard and {Brechignac}, Philippe and {Br{\"u}nken}, Sandra and {Buragohain}, Mridusmita and {Burkhardt}, Andrew and {Candian}, Alessandra and {Cazaux}, St{\'e}phanie and {Cernicharo}, Jose and {Chabot}, Marin and {Chakraborty}, Shubhadip and {Champion}, Jason and {Colgan}, Sean W.~J. and {Cooke}, Ilsa R. and {Coutens}, Audrey and {Cox}, Nick L.~J. and {Demyk}, Karine and {Meyer}, Jennifer Donovan and {Foschino}, Sacha and {Garc{\'\i}a-Lario}, Pedro and {Gavilan}, Lisseth and {Gerin}, Maryvonne and {Gottlieb}, Carl A. and {Guillard}, Pierre and {Gusdorf}, Antoine and {Hartigan}, Patrick and {He}, Jinhua and {Herbst}, Eric and {Hornekaer}, Liv and {J{\"a}ger}, Cornelia and {Janot-Pacheco}, Eduardo and {Kaufman}, Michael and {Kemper}, Francisca and {Kendrew}, Sarah and {Kirsanova}, Maria S. and {Klaassen}, Pamela and {Kwok}, Sun and {Labiano}, {\'A}lvaro and {Lai}, Thomas S.-Y. and {Lee}, Timothy J. and {Lefloch}, Bertrand and {Le Petit}, Franck and {Li}, Aigen and {Linz}, Hendrik and {Mackie}, Cameron J. and {Madden}, Suzanne C. and {Mascetti}, Jo{\"e}lle and {McGuire}, Brett A. and {Merino}, Pablo and {Micelotta}, Elisabetta R. and {Misselt}, Karl and {Morse}, Jon A. and {Mulas}, Giacomo and {Neelamkodan}, Naslim and {Ohsawa}, Ryou and {Omont}, Alain and {Paladini}, Roberta and {Palumbo}, Maria Elisabetta and {Pathak}, Amit and {Pendleton}, Yvonne J. and {Petrignani}, Annemieke and {Pino}, Thomas and {Puga}, Elena and {Rangwala}, Naseem and {Rapacioli}, Mathias and {Ricca}, Alessandra and {Roman-Duval}, Julia and {Roser}, Joseph and {Roueff}, Evelyne and {Rouill{\'e}}, Ga{\"e}l and {Salama}, Farid and {Sales}, Dinalva A. and {Sandstrom}, Karin and {Sarre}, Peter and {Sciamma-O'Brien}, Ella and {Sellgren}, Kris and {Shenoy}, Sachindev S. and {Teyssier}, David and {Thomas}, Richard D. and {Togi}, Aditya and {Verstraete}, Laurent and {Witt}, Adolf N. and {Wootten}, Alwyn and {Zettergren}, Henning and {Zhang}, Yong and {Zhang}, Ziwei E. and {Zhen}, Junfeng},
        title = "{PDRs4All. II. JWST's NIR and MIR imaging view of the Orion Nebula}",
      journal = {\aap},
         year = 2024,
        month = may,
       volume = {685},
          eid = {A73},
        pages = {A73},
          doi = {10.1051/0004-6361/202346747},
archivePrefix = {arXiv},
       eprint = {2308.16732},
 primaryClass = {astro-ph.GA},
       adsurl = {https://ui.adsabs.harvard.edu/abs/2024A&A...685A..73H}
}

@ARTICLE{Wolfire2022,
       author = {{Wolfire}, Mark G. and {Vallini}, Livia and {Chevance}, M{\'e}lanie},
        title = "{Photodissociation and X-Ray-Dominated Regions}",
      journal = {\araa},
         year = 2022,
        month = aug,
       volume = {60},
        pages = {247-318},
          doi = {10.1146/annurev-astro-052920-010254},
archivePrefix = {arXiv},
       eprint = {2202.05867},
 primaryClass = {astro-ph.GA},
       adsurl = {https://ui.adsabs.harvard.edu/abs/2022ARA&A..60..247W}
}

@ARTICLE{Berne2022,
       author = {{Bern{\'e}}, Olivier and {Habart}, {\'E}milie and {Peeters}, Els and {Abergel}, Alain and {Bergin}, Edwin A. and {Bernard-Salas}, Jeronimo and {Bron}, Emeric and {Cami}, Jan and {Dartois}, Emmanuel and {Fuente}, Asunci{\'o}n and {Goicoechea}, Javier R. and {Gordon}, Karl D. and {Okada}, Yoko and {Onaka}, Takashi and {Robberto}, Massimo and {R{\"o}llig}, Markus and {Tielens}, Alexander G.~G.~M. and {Vicente}, S{\'\i}lvia and {Wolfire}, Mark G. and {Alarc{\'o}n}, Felipe and {Boersma}, C. and {Canin}, Am{\'e}lie and {Chown}, Ryan and {Dicken}, Daniel and {Languignon}, David and {Le Gal}, Romane and {Pound}, Marc W. and {Trahin}, Boris and {Simmer}, Thomas and {Sidhu}, Ameek and {Van De Putte}, Dries and {Cuadrado}, Sara and {Guilloteau}, Claire and {Maragkoudakis}, Alexandros and {Schefter}, Bethany R. and {Schirmer}, Thi{\'e}baut and {Cazaux}, St{\'e}phanie and {Aleman}, Isabel and {Allamandola}, Louis and {Auchettl}, Rebecca and {Baratta}, Giuseppe Antonio and {Bejaoui}, Salma and {Bera}, Partha P. and {Bilalbegovi{\'c}}, Goranka and {Black}, John H. and {Boulanger}, Francois and {Bouwman}, Jordy and {Brandl}, Bernhard and {Brechignac}, Philippe and {Br{\"u}nken}, Sandra and {Burkhardt}, Andrew and {Candian}, Alessandra and {Cernicharo}, Jose and {Chabot}, Marin and {Chakraborty}, Shubhadip and {Champion}, Jason and {Colgan}, Sean W.~J. and {Cooke}, Ilsa R. and {Coutens}, Audrey and {Cox}, Nick L.~J. and {Demyk}, Karine and {Donovan Meyer}, Jennifer and {Engrand}, C{\'e}cile and {Foschino}, Sacha and {Garc{\'\i}a-Lario}, Pedro and {Gavilan}, Lisseth and {Gerin}, Maryvonne and {Godard}, Marie and {Gottlieb}, Carl A. and {Guillard}, Pierre and {Gusdorf}, Antoine and {Hartigan}, Patrick and {He}, Jinhua and {Herbst}, Eric and {Hornekaer}, Liv and {J{\"a}ger}, Cornelia and {Janot-Pacheco}, Eduardo and {Joblin}, Christine and {Kaufman}, Michael and {Kemper}, Francisca and {Kendrew}, Sarah and {Kirsanova}, Maria S. and {Klaassen}, Pamela and {Knight}, Collin and {Kwok}, Sun and {Labiano}, {\'A}lvaro and {Lai}, Thomas S.-Y. and {Lee}, Timothy J. and {Lefloch}, Bertrand and {Le Petit}, Franck and {Li}, Aigen and {Linz}, Hendrik and {Mackie}, Cameron J. and {Madden}, Suzanne C. and {Mascetti}, Jo{\"e}lle and {McGuire}, Brett A. and {Merino}, Pablo and {Micelotta}, Elisabetta R. and {Misselt}, Karl and {Morse}, Jon A. and {Mulas}, Giacomo and {Neelamkodan}, Naslim and {Ohsawa}, Ryou and {Omont}, Alain and {Paladini}, Roberta and {Palumbo}, Maria Elisabetta and {Pathak}, Amit and {Pendleton}, Yvonne J. and {Petrignani}, Annemieke and {Pino}, Thomas and {Puga}, Elena and {Rangwala}, Naseem and {Rapacioli}, Mathias and {Ricca}, Alessandra and {Roman-Duval}, Julia and {Roser}, Joseph and {Roueff}, Evelyne and {Rouill{\'e}}, Ga{\"e}l and {Salama}, Farid and {Sales}, Dinalva A. and {Sandstrom}, Karin and {Sarre}, Peter and {Sciamma-O'Brien}, Ella and {Sellgren}, Kris and {Shannon}, Matthew J. and {Shenoy}, Sachindev S. and {Teyssier}, David and {Thomas}, Richard D. and {Togi}, Aditya and {Verstraete}, Laurent and {Witt}, Adolf N. and {Wootten}, Alwyn and {Ysard}, Nathalie and {Zettergren}, Henning and {Zhang}, Yong and {Zhang}, Ziwei E. and {Zhen}, Junfeng},
        title = "{PDRs4All: A JWST Early Release Science Program on Radiative Feedback from Massive Stars}",
      journal = {\pasp},
         year = 2022,
        month = may,
       volume = {134},
       number = {1035},
          eid = {054301},
        pages = {054301},
          doi = {10.1088/1538-3873/ac604c},
archivePrefix = {arXiv},
       eprint = {2201.05112},
 primaryClass = {astro-ph.GA},
       adsurl = {https://ui.adsabs.harvard.edu/abs/2022PASP..134e4301B}
}

@ARTICLE{Zhang2021,
       author = {{Zhang}, Ziwei E. and {Cummings}, Sally J. and {Wan}, Yier and {Yang}, Benhui and {Stancil}, P.~C.},
        title = "{Properties of Highly Rotationally Excited H$_{2}$ in Photodissociation Regions}",
      journal = {\apj},
         year = 2021,
        month = may,
       volume = {912},
       number = {2},
          eid = {116},
        pages = {116},
          doi = {10.3847/1538-4357/abe9b0},
       adsurl = {https://ui.adsabs.harvard.edu/abs/2021ApJ...912..116Z}
}

@ARTICLE{Beattie2019,
       author = {{Beattie}, James R. and {Federrath}, Christoph and {Klessen}, Ralf S. and {Schneider}, Nicola},
        title = "{The relation between the turbulent Mach number and observed fractal dimensions of turbulent clouds}",
      journal = {\mnras},
         year = 2019,
        month = sep,
       volume = {488},
       number = {2},
        pages = {2493-2502},
          doi = {10.1093/mnras/stz1853},
archivePrefix = {arXiv},
       eprint = {1907.01689},
 primaryClass = {astro-ph.GA},
       adsurl = {https://ui.adsabs.harvard.edu/abs/2019MNRAS.488.2493B}
}

@ARTICLE{Kirsanova2019,
       author = {{Kirsanova}, Maria S. and {Wiebe}, Dmitri S.},
        title = "{Merged H/H$_{2}$ and C$^{+}$/C/CO transitions in the Orion Bar}",
      journal = {\mnras},
         year = 2019,
        month = jun,
       volume = {486},
       number = {2},
        pages = {2525-2534},
          doi = {10.1093/mnras/stz983},
archivePrefix = {arXiv},
       eprint = {1904.04423},
 primaryClass = {astro-ph.GA},
       adsurl = {https://ui.adsabs.harvard.edu/abs/2019MNRAS.486.2525K}
}

@ARTICLE{Cuadrado2019,
       author = {{Cuadrado}, S. and {Salas}, P. and {Goicoechea}, J.~R. and {Cernicharo}, J. and {Tielens}, A.~G.~G.~M. and {B{\'a}ez-Rubio}, A.},
        title = "{Direct estimation of electron density in the Orion Bar PDR from mm-wave carbon recombination lines}",
      journal = {\aap},
         year = 2019,
        month = may,
       volume = {625},
          eid = {L3},
        pages = {L3},
          doi = {10.1051/0004-6361/201935556},
archivePrefix = {arXiv},
       eprint = {1904.10356},
 primaryClass = {astro-ph.GA},
       adsurl = {https://ui.adsabs.harvard.edu/abs/2019A&A...625L...3C}
}

@ARTICLE{Goicoechea2019,
       author = {{Goicoechea}, Javier R. and {Santa-Maria}, Miriam G. and {Bron}, Emeric and {Teyssier}, David and {Marcelino}, Nuria and {Cernicharo}, Jos{\'e} and {Cuadrado}, Sara},
        title = "{Molecular tracers of radiative feedback in Orion (OMC-1). Widespread CH$^{+}$ (J = 1-0), CO (10-9), HCN (6-5), and HCO$^{+}$ (6-5) emission}",
      journal = {\aap},
         year = 2019,
        month = feb,
       volume = {622},
          eid = {A91},
        pages = {A91},
          doi = {10.1051/0004-6361/201834409},
archivePrefix = {arXiv},
       eprint = {1812.00821},
 primaryClass = {astro-ph.GA},
       adsurl = {https://ui.adsabs.harvard.edu/abs/2019A&A...622A..91G}
}

@ARTICLE{Joblin2018,
       author = {{Joblin}, C. and {Bron}, E. and {Pinto}, C. and {Pilleri}, P. and {Le Petit}, F. and {Gerin}, M. and {Le Bourlot}, J. and {Fuente}, A. and {Berne}, O. and {Goicoechea}, J.~R. and {Habart}, E. and {K{\"o}hler}, M. and {Teyssier}, D. and {Nagy}, Z. and {Montillaud}, J. and {Vastel}, C. and {Cernicharo}, J. and {R{\"o}llig}, M. and {Ossenkopf-Okada}, V. and {Bergin}, E.~A.},
        title = "{Structure of photodissociation fronts in star-forming regions revealed by Herschel observations of high-J CO emission lines}",
      journal = {\aap},
         year = 2018,
        month = jul,
       volume = {615},
          eid = {A129},
        pages = {A129},
          doi = {10.1051/0004-6361/201832611},
archivePrefix = {arXiv},
       eprint = {1801.03893},
 primaryClass = {astro-ph.GA},
       adsurl = {https://ui.adsabs.harvard.edu/abs/2018A&A...615A.129J}
}

@ARTICLE{Bron2018,
       author = {{Bron}, Emeric and {Ag{\'u}ndez}, Marcelino and {Goicoechea}, Javier R. and {Cernicharo}, Jos{\'e}},
        title = "{Photoevaporating PDR models with the Hydra PDR Code}",
      journal = {arXiv e-prints},
         year = 2018,
        month = jan,
          eid = {arXiv:1801.01547},
        pages = {arXiv:1801.01547},
          doi = {10.48550/arXiv.1801.01547},
archivePrefix = {arXiv},
       eprint = {1801.01547},
 primaryClass = {astro-ph.GA},
       adsurl = {https://ui.adsabs.harvard.edu/abs/2018arXiv180101547B}
}

@ARTICLE{Andree-Labsch2017,
       author = {{Andree-Labsch}, S. and {Ossenkopf-Okada}, V. and {R{\"o}llig}, M.},
        title = "{Modelling clumpy photon-dominated regions in 3D. Understanding the Orion Bar stratification}",
      journal = {\aap},
         year = 2017,
        month = feb,
       volume = {598},
          eid = {A2},
        pages = {A2},
          doi = {10.1051/0004-6361/201424287},
archivePrefix = {arXiv},
       eprint = {1405.5553},
 primaryClass = {astro-ph.SR},
       adsurl = {https://ui.adsabs.harvard.edu/abs/2017A&A...598A...2A}
}

@ARTICLE{Goicoechea2016,
       author = {{Goicoechea}, Javier R. and {Pety}, J{\'e}r{\^o}me and {Cuadrado}, Sara and {Cernicharo}, Jos{\'e} and {Chapillon}, Edwige and {Fuente}, Asunci{\'o}n and {Gerin}, Maryvonne and {Joblin}, Christine and {Marcelino}, Nuria and {Pilleri}, Paolo},
        title = "{Compression and ablation of the photo-irradiated molecular cloud the Orion Bar}",
      journal = {\nat},
         year = 2016,
        month = sep,
       volume = {537},
       number = {7619},
        pages = {207-209},
          doi = {10.1038/nature18957},
archivePrefix = {arXiv},
       eprint = {1608.06173},
 primaryClass = {astro-ph.GA},
       adsurl = {https://ui.adsabs.harvard.edu/abs/2016Natur.537..207G}
}

@ARTICLE{Cuadrado2015,
       author = {{Cuadrado}, S. and {Goicoechea}, J.~R. and {Pilleri}, P. and {Cernicharo}, J. and {Fuente}, A. and {Joblin}, C.},
        title = "{The chemistry and spatial distribution of small hydrocarbons in UV-irradiated molecular clouds: the Orion Bar PDR}",
      journal = {\aap},
         year = 2015,
        month = mar,
       volume = {575},
          eid = {A82},
        pages = {A82},
          doi = {10.1051/0004-6361/201424568},
archivePrefix = {arXiv},
       eprint = {1412.0417},
 primaryClass = {astro-ph.GA},
       adsurl = {https://ui.adsabs.harvard.edu/abs/2015A&A...575A..82C}
}

@ARTICLE{Nagy2013,
       author = {{Nagy}, Z. and {Van der Tak}, F.~F.~S. and {Ossenkopf}, V. and {Gerin}, M. and {Le Petit}, F. and {Le Bourlot}, J. and {Black}, J.~H. and {Goicoechea}, J.~R. and {Joblin}, C. and {R{\"o}llig}, M. and {Bergin}, E.~A.},
        title = "{The chemistry of ions in the Orion Bar I. - CH$^{+}$, SH$^{+}$, and CF$^{+}$. The effect of high electron density and vibrationally excited H$_{2}$ in a warm PDR surface}",
      journal = {\aap},
         year = 2013,
        month = feb,
       volume = {550},
          eid = {A96},
        pages = {A96},
          doi = {10.1051/0004-6361/201220519},
archivePrefix = {arXiv},
       eprint = {1212.4378},
 primaryClass = {astro-ph.GA},
       adsurl = {https://ui.adsabs.harvard.edu/abs/2013A&A...550A..96N}
}

@ARTICLE{Bisbas2012,
       author = {{Bisbas}, T.~G. and {Bell}, T.~A. and {Viti}, S. and {Yates}, J. and {Barlow}, M.~J.},
        title = "{3D-PDR: a new three-dimensional astrochemistry code for treating photodissociation regions}",
      journal = {\mnras},
         year = 2012,
        month = dec,
       volume = {427},
       number = {3},
        pages = {2100-2118},
          doi = {10.1111/j.1365-2966.2012.22077.x},
archivePrefix = {arXiv},
       eprint = {1209.1091},
 primaryClass = {astro-ph.SR},
       adsurl = {https://ui.adsabs.harvard.edu/abs/2012MNRAS.427.2100B}
}

@ARTICLE{Walch2012,
       author = {{Walch}, S.~K. and {Whitworth}, A.~P. and {Bisbas}, T. and {W{\"u}nsch}, R. and {Hubber}, D.},
        title = "{Dispersal of molecular clouds by ionizing radiation}",
      journal = {\mnras},
         year = 2012,
        month = nov,
       volume = {427},
       number = {1},
        pages = {625-636},
          doi = {10.1111/j.1365-2966.2012.21767.x},
archivePrefix = {arXiv},
       eprint = {1206.6492},
 primaryClass = {astro-ph.GA},
       adsurl = {https://ui.adsabs.harvard.edu/abs/2012MNRAS.427..625W}
}

@ARTICLE{Agundez2010,
       author = {{Ag{\'u}ndez}, M. and {Goicoechea}, J.~R. and {Cernicharo}, J. and {Faure}, A. and {Roueff}, E.},
        title = "{The Chemistry of Vibrationally Excited H$_{2}$ in the Interstellar Medium}",
      journal = {\apj},
         year = 2010,
        month = apr,
       volume = {713},
       number = {1},
        pages = {662-670},
          doi = {10.1088/0004-637X/713/1/662},
archivePrefix = {arXiv},
       eprint = {1003.1375},
 primaryClass = {astro-ph.GA},
       adsurl = {https://ui.adsabs.harvard.edu/abs/2010ApJ...713..662A}
}

@ARTICLE{LePetit2006,
       author = {{Le Petit}, Franck and {Nehm{\'e}}, Cyrine and {Le Bourlot}, Jacques and {Roueff}, Evelyne},
        title = "{A Model for Atomic and Molecular Interstellar Gas: The Meudon PDR Code}",
      journal = {\apjs},
         year = 2006,
        month = jun,
       volume = {164},
       number = {2},
        pages = {506-529},
          doi = {10.1086/503252},
archivePrefix = {arXiv},
       eprint = {astro-ph/0602150},
 primaryClass = {astro-ph},
       adsurl = {https://ui.adsabs.harvard.edu/abs/2006ApJS..164..506L}
}

@ARTICLE{Sofia2004,
       author = {{Sofia}, Ulysses J. and {Lauroesch}, James T. and {Meyer}, David M. and {Cartledge}, Stefan I.~B.},
        title = "{Interstellar Carbon in Translucent Sight Lines}",
      journal = {\apj},
         year = 2004,
        month = apr,
       volume = {605},
       number = {1},
        pages = {272-277},
          doi = {10.1086/382592},
archivePrefix = {arXiv},
       eprint = {astro-ph/0401510},
 primaryClass = {astro-ph},
       adsurl = {https://ui.adsabs.harvard.edu/abs/2004ApJ...605..272S}
}

@ARTICLE{Walmsley2000,
       author = {{Walmsley}, C.~M. and {Natta}, A. and {Oliva}, E. and {Testi}, L.},
        title = "{The structure of the Orion bar}",
      journal = {\aap},
         year = 2000,
        month = dec,
       volume = {364},
        pages = {301-317},
       adsurl = {https://ui.adsabs.harvard.edu/abs/2000A&A...364..301W}
}

@ARTICLE{Stutzki1998,
       author = {{Stutzki}, J. and {Bensch}, F. and {Heithausen}, A. and {Ossenkopf}, V. and {Zielinsky}, M.},
        title = "{On the fractal structure of molecular clouds}",
      journal = {\aap},
         year = 1998,
        month = aug,
       volume = {336},
        pages = {697-720},
       adsurl = {https://ui.adsabs.harvard.edu/abs/1998A&A...336..697S}
}

@ARTICLE{Meyer1997,
       author = {{Meyer}, David M. and {Cardelli}, Jason A. and {Sofia}, Ulysses J.},
        title = "{The Abundance of Interstellar Nitrogen}",
      journal = {\apjl},
         year = 1997,
        month = nov,
       volume = {490},
       number = {1},
        pages = {L103-L106},
          doi = {10.1086/311023},
archivePrefix = {arXiv},
       eprint = {astro-ph/9710162},
 primaryClass = {astro-ph},
       adsurl = {https://ui.adsabs.harvard.edu/abs/1997ApJ...490L.103M}
}

@ARTICLE{Wyrowski1997,
       author = {{Wyrowski}, F. and {Schilke}, P. and {Hofner}, P. and {Walmsley}, C.~M.},
        title = "{Carbon Radio Recombination Lines in the Orion Bar}",
      journal = {\apjl},
         year = 1997,
        month = oct,
       volume = {487},
       number = {2},
        pages = {L171-L174},
          doi = {10.1086/310893},
archivePrefix = {arXiv},
       eprint = {astro-ph/9707276},
 primaryClass = {astro-ph},
       adsurl = {https://ui.adsabs.harvard.edu/abs/1997ApJ...487L.171W}
}

@ARTICLE{Hollenbach1997,
       author = {{Hollenbach}, D.~J. and {Tielens}, A.~G.~G.~M.},
        title = "{Dense Photodissociation Regions (PDRs)}",
      journal = {\araa},
         year = 1997,
        month = jan,
       volume = {35},
        pages = {179-216},
          doi = {10.1146/annurev.astro.35.1.179},
       adsurl = {https://ui.adsabs.harvard.edu/abs/1997ARA&A..35..179H}
}

@ARTICLE{Elmegreen1996,
       author = {{Elmegreen}, Bruce G. and {Falgarone}, Edith},
        title = "{A Fractal Origin for the Mass Spectrum of Interstellar Clouds}",
      journal = {\apj},
         year = 1996,
        month = nov,
       volume = {471},
        pages = {816},
          doi = {10.1086/178009},
       adsurl = {https://ui.adsabs.harvard.edu/abs/1996ApJ...471..816E}
}

@ARTICLE{Bertoldi1996,
       author = {{Bertoldi}, Frank and {Draine}, B.~T.},
        title = "{Nonequilibrium Photodissociation Regions: Ionization-Dissociation Fronts}",
      journal = {\apj},
         year = 1996,
        month = feb,
       volume = {458},
        pages = {222},
          doi = {10.1086/176805},
archivePrefix = {arXiv},
       eprint = {astro-ph/9508067},
 primaryClass = {astro-ph},
       adsurl = {https://ui.adsabs.harvard.edu/abs/1996ApJ...458..222B}
}

@ARTICLE{Tielens1993,
       author = {{Tielens}, A.~G.~G.~M. and {Meixner}, M.~M. and {van der Werf}, P.~P. and {Bregman}, J. and {Tauber}, J.~A. and {Stutzki}, J. and {Rank}, D.},
        title = "{Anatomy of the Photodissociation Region in the Orion Bar}",
      journal = {Science},
         year = 1993,
        month = oct,
       volume = {262},
       number = {5130},
        pages = {86-89},
          doi = {10.1126/science.262.5130.86},
       adsurl = {https://ui.adsabs.harvard.edu/abs/1993Sci...262...86T}
}

@INPROCEEDINGS{Scalo1990,
       author = {{Scalo}, John},
        title = "{Perception of interstellar structure - Facing complexity}",
    booktitle = {Physical Processes in Fragmentation and Star Formation},
         year = 1990,
       editor = {{Capuzzo-Dolcetta}, Roberto and {Chiosi}, Cesare and {di Fazio}, Alberto},
       series = {Astrophysics and Space Science Library},
       volume = {162},
        month = jan,
        pages = {151-176},
          doi = {10.1007/978-94-009-0605-1_12},
       adsurl = {https://ui.adsabs.harvard.edu/abs/1990ASSL..162..151S}
}

@ARTICLE{Tielens1985,
       author = {{Tielens}, A.~G.~G.~M. and {Hollenbach}, D.},
        title = "{Photodissociation regions. I. Basic model.}",
      journal = {\apj},
         year = 1985,
        month = apr,
       volume = {291},
        pages = {722-746},
          doi = {10.1086/163111},
       adsurl = {https://ui.adsabs.harvard.edu/abs/1985ApJ...291..722T}
}

@ARTICLE{Draine1978,
       author = {{Draine}, B.~T.},
        title = "{Photoelectric heating of interstellar gas.}",
      journal = {\apjs},
         year = 1978,
        month = apr,
       volume = {36},
        pages = {595-619},
          doi = {10.1086/190513},
       adsurl = {https://ui.adsabs.harvard.edu/abs/1978ApJS...36..595D}
}

@ARTICLE{LeBourlot2000,
       author = {{Le Bourlot}, J.},
        title = "{Ortho to para conversion of H$_{2}$ on interstellar grains}",
      journal = {\aap},
         year = 2000,
        month = aug,
       volume = {360},
        pages = {656-662},
       adsurl = {https://ui.adsabs.harvard.edu/abs/2000A&A...360..656L}
}

@ARTICLE{Zannese2025b,
       author = {{Zannese}, M. and {Tabone}, B. and {Habart}, E. and {Dartois}, E. and {Goicoechea}, J.~R. and {Coudert}, L. and {Gans}, B. and {Martin-Drumel}, M.-A. and {Jacovella}, U. and {Faure}, A. and {Godard}, B. and {Tielens}, A.~G.~G.~M. and {Le Gal}, R. and {Black}, J.~H. and {Vicente}, S. and {Bern{\'e}}, O. and {Peeters}, E. and {Van De Putte}, D. and {Chown}, R. and {Sidhu}, A. and {Schroetter}, I. and {Canin}, A. and {Kannavou}, O.},
        title = "{PDRs4All: XI. Detection of infrared CH$^{+}$ and CH$_{3}$$^{+}$ rovibrational emission in the Orion Bar and disk d203-506: Evidence of chemical pumping}",
      journal = {\aap},
         year = 2025,
        month = apr,
       volume = {696},
          eid = {A99},
        pages = {A99},
          doi = {10.1051/0004-6361/202453441},
archivePrefix = {arXiv},
       eprint = {2502.08354},
 primaryClass = {astro-ph.GA},
       adsurl = {https://ui.adsabs.harvard.edu/abs/2025A&A...696A..99Z}
}

@ARTICLE{Zanchet2013,
       author = {{Zanchet}, Alexandre and {Godard}, B. and {Bulut}, Niyazi and {Roncero}, Octavio and {Halvick}, Philippe and {Cernicharo}, Jos{\'e}},
        title = "{H$_{2}$(v = 0,1) + C$^{+}$($^{2}$ P) {\textrightarrow} H+CH$^{+}$ State-to-state Rate Constants for Chemical Pumping Models in Astrophysical Media}",
      journal = {\apj},
         year = 2013,
        month = apr,
       volume = {766},
       number = {2},
          eid = {80},
        pages = {80},
          doi = {10.1088/0004-637X/766/2/80},
       adsurl = {https://ui.adsabs.harvard.edu/abs/2013ApJ...766...80Z}
}

@ARTICLE{Abgrall2000,
       author = {{Abgrall}, H. and {Roueff}, E. and {Drira}, I.},
        title = "{Total transition probability and spontaneous radiative dissociation of B, C, B' and D states of molecular hydrogen}",
      journal = {\aaps},
         year = 2000,
        month = jan,
       volume = {141},
        pages = {297-300},
          doi = {10.1051/aas:2000121},
       adsurl = {https://ui.adsabs.harvard.edu/abs/2000A&AS..141..297A}
}

@ARTICLE{Herraez-Aguilar2014,
       author = {{Herr{\'a}ez-Aguilar}, D. and {Jambrina}, P.~G. and {Men{\'e}ndez}, M. and {Aldegunde}, J. and {Warmbier}, R. and {Aoiz}, F.~J.},
        title = "{The effect of the reactant internal excitation on the dynamics of the C++ H2reaction}",
      journal = {Physical Chemistry Chemical Physics (Incorporating Faraday Transactions)},
         year = 2014,
        month = jan,
       volume = {16},
       number = {45},
        pages = {24800-24812},
          doi = {10.1039/C4CP03289F},
       adsurl = {https://ui.adsabs.harvard.edu/abs/2014PCCP...1624800H}
}

@ARTICLE{Sternberg1999,
       author = {{Sternberg}, Amiel and {Neufeld}, David A.},
        title = "{The Ratio of Ortho- to Para-H$_{2}$ in Photodissociation Regions}",
      journal = {\apj},
         year = 1999,
        month = may,
       volume = {516},
       number = {1},
        pages = {371-380},
          doi = {10.1086/307115},
archivePrefix = {arXiv},
       eprint = {astro-ph/9812049},
 primaryClass = {astro-ph},
       adsurl = {https://ui.adsabs.harvard.edu/abs/1999ApJ...516..371S}
}

@article{Komasa2011,
author = {Komasa, Jacek and Piszczatowski, Konrad and Łach, Grzegorz and Przybytek, Michał and Jeziorski, Bogumił and Pachucki, Krzysztof},
title = {Quantum Electrodynamics Effects in Rovibrational Spectra of Molecular Hydrogen},
journal = {Journal of Chemical Theory and Computation},
volume = {7},
number = {10},
pages = {3105-3115},
year = {2011},
doi = {10.1021/ct200438t},
note ={PMID: 26598154},
URL = {   
        https://doi.org/10.1021/ct200438t
},
eprint = {   
        https://doi.org/10.1021/ct200438t
}
}

@ARTICLE{Shaw2005,
       author = {{Shaw}, G. and {Ferland}, G.~J. and {Abel}, N.~P. and {Stancil}, P.~C. and {van Hoof}, P.~A.~M.},
        title = "{Molecular Hydrogen in Star-forming Regions: Implementation of its Microphysics in CLOUDY}",
      journal = {\apj},
         year = 2005,
        month = may,
       volume = {624},
       number = {2},
        pages = {794-807},
          doi = {10.1086/429215},
archivePrefix = {arXiv},
       eprint = {astro-ph/0501485},
 primaryClass = {astro-ph},
       adsurl = {https://ui.adsabs.harvard.edu/abs/2005ApJ...624..794S}
}

@ARTICLE{Draine1996,
       author = {{Draine}, B.~T. and {Bertoldi}, Frank},
        title = "{Structure of Stationary Photodissociation Fronts}",
      journal = {\apj},
         year = 1996,
        month = sep,
       volume = {468},
        pages = {269},
          doi = {10.1086/177689},
archivePrefix = {arXiv},
       eprint = {astro-ph/9603032},
 primaryClass = {astro-ph},
       adsurl = {https://ui.adsabs.harvard.edu/abs/1996ApJ...468..269D}
}

@ARTICLE{Sternberg1989a,
       author = {{Sternberg}, Amiel},
        title = "{Ultraviolet Fluorescent Molecular Hydrogen Emission}",
      journal = {\apj},
         year = 1989,
        month = dec,
       volume = {347},
        pages = {863},
          doi = {10.1086/168177},
       adsurl = {https://ui.adsabs.harvard.edu/abs/1989ApJ...347..863S}
}

@ARTICLE{Sternberg1989b,
       author = {{Sternberg}, A. and {Dalgarno}, A.},
        title = "{The Infrared Response of Molecular Hydrogen Gas to Ultraviolet Radiation: High-Density Regions}",
      journal = {\apj},
         year = 1989,
        month = mar,
       volume = {338},
        pages = {197},
          doi = {10.1086/167193},
       adsurl = {https://ui.adsabs.harvard.edu/abs/1989ApJ...338..197S}
}

@ARTICLE{Black1987,
       author = {{Black}, John H. and {van Dishoeck}, Ewine F.},
        title = "{Fluorescent Excitation of Interstellar H 2}",
      journal = {\apj},
         year = 1987,
        month = nov,
       volume = {322},
        pages = {412},
          doi = {10.1086/165740},
       adsurl = {https://ui.adsabs.harvard.edu/abs/1987ApJ...322..412B}
}

@ARTICLE{vanDishoeck1986,
       author = {{van Dishoeck}, E.~F. and {Black}, J.~H.},
        title = "{Comprehensive Models of Diffuse Interstellar Clouds: Physical Conditions and Molecular Abundances}",
      journal = {\apjs},
         year = 1986,
        month = sep,
       volume = {62},
        pages = {109},
          doi = {10.1086/191135},
       adsurl = {https://ui.adsabs.harvard.edu/abs/1986ApJS...62..109V}
}

@ARTICLE{Rollig2022,
       author = {{R{\"o}llig}, M. and {Ossenkopf-Okada}, V.},
        title = "{The KOSMA-{\ensuremath{\tau}} PDR model. I. Recent updates to the numerical model of photo-dissociated regions}",
      journal = {\aap},
         year = 2022,
        month = aug,
       volume = {664},
          eid = {A67},
        pages = {A67},
          doi = {10.1051/0004-6361/202141854},
archivePrefix = {arXiv},
       eprint = {2205.04233},
 primaryClass = {astro-ph.GA},
       adsurl = {https://ui.adsabs.harvard.edu/abs/2022A&A...664A..67R}
}

@ARTICLE{Wolniewicz1998,
       author = {{Wolniewicz}, L. and {Simbotin}, I. and {Dalgarno}, A.},
        title = "{Quadrupole Transition Probabilities for the Excited Rovibrational States of H$_{2}$}",
      journal = {\apjs},
         year = 1998,
        month = apr,
       volume = {115},
       number = {2},
        pages = {293-313},
          doi = {10.1086/313091},
       adsurl = {https://ui.adsabs.harvard.edu/abs/1998ApJS..115..293W}
}

@ARTICLE{Habart2023,
       author = {{Habart}, Emilie and {Le Gal}, Romane and {Alvarez}, Carlos and {Peeters}, Els and {Bern{\'e}}, Olivier and {Wolfire}, Mark G. and {Goicoechea}, Javier R. and {Schirmer}, Thi{\'e}baut and {Bron}, Emeric and {R{\"o}llig}, Markus},
        title = "{High-angular-resolution NIR view of the Orion Bar revealed by Keck/NIRC2}",
      journal = {\aap},
         year = 2023,
        month = may,
       volume = {673},
          eid = {A149},
        pages = {A149},
          doi = {10.1051/0004-6361/202244034},
archivePrefix = {arXiv},
       eprint = {2206.08245},
 primaryClass = {astro-ph.GA},
       adsurl = {https://ui.adsabs.harvard.edu/abs/2023A&A...673A.149H}
}

@ARTICLE{Sanchez2007,
       author = {{S{\'a}nchez}, N{\'e}stor and {Alfaro}, Emilio J. and {P{\'e}rez}, Enrique},
        title = "{Fractal Dimension of Interstellar Clouds: Opacity and Noise Effects}",
      journal = {\apj},
         year = 2007,
        month = feb,
       volume = {656},
       number = {1},
        pages = {222-226},
          doi = {10.1086/510351},
archivePrefix = {arXiv},
       eprint = {astro-ph/0610613},
 primaryClass = {astro-ph},
       adsurl = {https://ui.adsabs.harvard.edu/abs/2007ApJ...656..222S}
}

@ARTICLE{Hunt2025,
       author = {{Hunt}, L.~K. and {Draine}, B.~T. and {Navarro}, M.~G. and {Aloisi}, A. and {Vaught}, R.~J. Rickards and {Adamo}, A. and {Annibali}, F. and {Calzetti}, D. and {Hernandez}, S. and {James}, B.~L. and {Mingozzi}, M. and {Schneider}, R. and {Tosi}, M. and {Brandl}, B. and {del Valle-Espinosa}, M.~G. and {Donnan}, F. and {Hirschauer}, A.~S. and {Meixner}, M. and {Rigopoulou}, D.},
        title = "{The Interstellar Medium in I Zw 18 Seen with JWST/MIRI. II. Warm Molecular Hydrogen and Warm Dust}",
      journal = {\apj},
         year = 2025,
        month = nov,
       volume = {993},
       number = {1},
          eid = {84},
        pages = {84},
          doi = {10.3847/1538-4357/ae0191},
archivePrefix = {arXiv},
       eprint = {2509.02690},
 primaryClass = {astro-ph.GA},
       adsurl = {https://ui.adsabs.harvard.edu/abs/2025ApJ...993...84H}
}

@article{pitts1982,
    title = {Temperature dependence of the C2(X1Σg+) reaction with H2 and CH4 and C2(X1Σg+ and a 3Πu equilibrated states) with O2},
    journal = {Chemical Physics},
    volume = {68},
    number = {3},
    pages = {417-422},
    year = {1982},
    issn = {0301-0104},
    doi = {https://doi.org/10.1016/0301-0104(82)87050-X},
    url = {https://www.sciencedirect.com/science/article/pii/030101048287050X},
    author = {Willam M. Pitts and L. Pasternack and J.R. McDonald}
}

@ARTICLE{Flannery1980,
       author = {{Flannery}, B.~P. and {Roberge}, W. and {Rybicki}, G.~B.},
        title = "{The penetration of diffuse ultraviolet radiation into interstellar clouds}",
      journal = {\apj},
         year = 1980,
        month = mar,
       volume = {236},
        pages = {598-608},
          doi = {10.1086/157778},
       adsurl = {https://ui.adsabs.harvard.edu/abs/1980ApJ...236..598F}
}

@ARTICLE{Goicoechea2007,
       author = {{Goicoechea}, J.~R. and {Le Bourlot}, J.},
        title = "{The penetration of Far-UV radiation into molecular clouds}",
      journal = {\aap},
         year = 2007,
        month = may,
       volume = {467},
       number = {1},
        pages = {1-14},
          doi = {10.1051/0004-6361:20066119},
archivePrefix = {arXiv},
       eprint = {astro-ph/0702033},
 primaryClass = {astro-ph},
       adsurl = {https://ui.adsabs.harvard.edu/abs/2007A&A...467....1G}
}

@ARTICLE{Zanchet2019,
       author = {{Zanchet}, Alexandre and {Lique}, Fran{\c{c}}ois and {Roncero}, Octavio and {Goicoechea}, Javier R. and {Bulut}, Niyazi},
        title = "{Formation of interstellar SH$^{+}$ from vibrationally excited H$_{2}$: Quantum study of S$^{+}$ + H$_{2}$ ⇄ SH$^{+}$ + H reaction and inelastic collision}",
      journal = {\aap},
         year = 2019,
        month = jun,
       volume = {626},
          eid = {A103},
        pages = {A103},
          doi = {10.1051/0004-6361/201935471},
archivePrefix = {arXiv},
       eprint = {1905.02779},
 primaryClass = {astro-ph.GA},
       adsurl = {https://ui.adsabs.harvard.edu/abs/2019A&A...626A.103Z}
}

@ARTICLE{Millar2024,
       author = {{Millar}, T.~J. and {Walsh}, C. and {Van de Sande}, M. and {Markwick}, A.~J.},
        title = "{The UMIST Database for Astrochemistry 2022}",
      journal = {\aap},
         year = 2024,
        month = feb,
       volume = {682},
          eid = {A109},
        pages = {A109},
          doi = {10.1051/0004-6361/202346908},
archivePrefix = {arXiv},
       eprint = {2311.03936},
 primaryClass = {astro-ph.GA},
       adsurl = {https://ui.adsabs.harvard.edu/abs/2024A&A...682A.109M}
}

@ARTICLE{Wakelam2024,
       author = {{Wakelam}, V. and {Gratier}, P. and {Loison}, J.-C. and {Hickson}, K.~M. and {Penguen}, J. and {Mechineau}, A.},
        title = "{The 2024 KIDA network for interstellar chemistry}",
      journal = {\aap},
         year = 2024,
        month = sep,
       volume = {689},
          eid = {A63},
        pages = {A63},
          doi = {10.1051/0004-6361/202450606},
archivePrefix = {arXiv},
       eprint = {2407.15958},
 primaryClass = {astro-ph.GA},
       adsurl = {https://ui.adsabs.harvard.edu/abs/2024A&A...689A..63W}
}

@ARTICLE{vanDishoeck1988,
       author = {{van Dishoeck}, Ewine F. and {Black}, John H.},
        title = "{The Photodissociation and Chemistry of Interstellar CO}",
      journal = {\apj},
         year = 1988,
        month = nov,
       volume = {334},
        pages = {771},
          doi = {10.1086/166877},
       adsurl = {https://ui.adsabs.harvard.edu/abs/1988ApJ...334..771V}
}

@ARTICLE{Federman1979,
       author = {{Federman}, S.~R. and {Glassgold}, A.~E. and {Kwan}, J.},
        title = "{Atomic to molecular hydrogen transition in interstellar clouds.}",
      journal = {\apj},
         year = 1979,
        month = jan,
       volume = {227},
        pages = {466-473},
          doi = {10.1086/156753},
       adsurl = {https://ui.adsabs.harvard.edu/abs/1979ApJ...227..466F}
}

@ARTICLE{Cazaux2002,
       author = {{Cazaux}, S. and {Tielens}, A.~G.~G.~M.},
        title = "{Molecular Hydrogen Formation in the Interstellar Medium}",
      journal = {\apjl},
         year = 2002,
        month = aug,
       volume = {575},
       number = {1},
        pages = {L29-L32},
          doi = {10.1086/342607},
archivePrefix = {arXiv},
       eprint = {astro-ph/0207035},
 primaryClass = {astro-ph},
       adsurl = {https://ui.adsabs.harvard.edu/abs/2002ApJ...575L..29C}
}

@ARTICLE{Hogerheijde1995,
       author = {{Hogerheijde}, Michiel R. and {Jansen}, David J. and {van Dishoeck}, Ewine F.},
        title = "{Millimeter and submillimeter observations of the Orion Bar. I. Physical structure.}",
      journal = {\aap},
         year = 1995,
        month = feb,
       volume = {294},
        pages = {792-810},
       adsurl = {https://ui.adsabs.harvard.edu/abs/1995A&A...294..792H}
}

@ARTICLE{Weingartner2001b,
       author = {{Weingartner}, Joseph C. and {Draine}, B.~T.},
        title = "{Electron-Ion Recombination on Grains and Polycyclic Aromatic Hydrocarbons}",
      journal = {\apj},
         year = 2001,
        month = dec,
       volume = {563},
       number = {2},
        pages = {842-852},
          doi = {10.1086/324035},
archivePrefix = {arXiv},
       eprint = {astro-ph/0105237},
 primaryClass = {astro-ph},
       adsurl = {https://ui.adsabs.harvard.edu/abs/2001ApJ...563..842W}
}

@ARTICLE{Draine2003,
       author = {{Draine}, B.~T.},
        title = "{Interstellar Dust Grains}",
      journal = {\araa},
         year = 2003,
        month = jan,
       volume = {41},
        pages = {241-289},
          doi = {10.1146/annurev.astro.41.011802.094840},
archivePrefix = {arXiv},
       eprint = {astro-ph/0304489},
 primaryClass = {astro-ph},
       adsurl = {https://ui.adsabs.harvard.edu/abs/2003ARA&A..41..241D}
}

@ARTICLE{Weingartner2001a,
       author = {{Weingartner}, Joseph C. and {Draine}, B.~T.},
        title = "{Dust Grain-Size Distributions and Extinction in the Milky Way, Large Magellanic Cloud, and Small Magellanic Cloud}",
      journal = {\apj},
         year = 2001,
        month = feb,
       volume = {548},
       number = {1},
        pages = {296-309},
          doi = {10.1086/318651},
archivePrefix = {arXiv},
       eprint = {astro-ph/0008146},
 primaryClass = {astro-ph},
       adsurl = {https://ui.adsabs.harvard.edu/abs/2001ApJ...548..296W}
}

@ARTICLE{Fuente2024,
       author = {{Fuente}, Asunci{\'o}n and {Roueff}, Evelyne and {Le Petit}, Franck and {Le Bourlot}, Jacques and {Bron}, Emeric and {Wolfire}, Mark G. and {Babb}, James F. and {Yan}, Pei-Gen and {Onaka}, Takashi and {Black}, John H. and {Schroetter}, Ilane and {Van De Putte}, Dries and {Sidhu}, Ameek and {Canin}, Am{\'e}lie and {Trahin}, Boris and {Alarc{\'o}n}, Felipe and {Chown}, Ryan and {Kannavou}, Olga and {Bern{\'e}}, Olivier and {Habart}, Emilie and {Peeters}, Els and {Goicoechea}, Javier R. and {Zannese}, Marion and {Meshaka}, Raphael and {Okada}, Yoko and {R{\"o}llig}, Markus and {Le Gal}, Romane and {Sales}, Dinalva A. and {Palumbo}, Maria Elisabetta and {Baratta}, Giuseppe Antonio and {Madden}, Suzanne C. and {Neelamkodan}, Naslim and {Zhang}, Ziwei E. and {Stancil}, P.~C.},
        title = "{PDRs4All. IX. Sulfur elemental abundance in the Orion Bar}",
      journal = {\aap},
         year = 2024,
        month = jul,
       volume = {687},
          eid = {A87},
        pages = {A87},
          doi = {10.1051/0004-6361/202449229},
archivePrefix = {arXiv},
       eprint = {2404.09235},
 primaryClass = {astro-ph.GA},
       adsurl = {https://ui.adsabs.harvard.edu/abs/2024A&A...687A..87F}
}

@BOOK{Sobolev1960,
       author = {{Sobolev}, V.~V.},
        title = "{Moving Envelopes of Stars}",
         year = 1960,
          doi = {10.4159/harvard.9780674864658},
       adsurl = {https://ui.adsabs.harvard.edu/abs/1960mes..book.....S}
}

@ARTICLE{deJong1975,
       author = {{de Jong}, T. and {Chu}, S. and {Dalgarno}, A.},
        title = "{Carbon monoxide in collapsing interstellar clouds.}",
      journal = {\apj},
         year = 1975,
        month = jul,
       volume = {199},
        pages = {69-78},
          doi = {10.1086/153665},
       adsurl = {https://ui.adsabs.harvard.edu/abs/1975ApJ...199...69D}
}

@ARTICLE{Berne2023,
       author = {{Bern{\'e}}, Olivier and {Martin-Drumel}, Marie-Aline and {Schroetter}, Ilane and {Goicoechea}, Javier R. and {Jacovella}, Ugo and {Gans}, B{\'e}renger and {Dartois}, Emmanuel and {Coudert}, Laurent H. and {Bergin}, Edwin and {Alarcon}, Felipe and {Cami}, Jan and {Roueff}, Evelyne and {Black}, John H. and {Asvany}, Oskar and {Habart}, Emilie and {Peeters}, Els and {Canin}, Amelie and {Trahin}, Boris and {Joblin}, Christine and {Schlemmer}, Stephan and {Thorwirth}, Sven and {Cernicharo}, Jose and {Gerin}, Maryvonne and {Tielens}, Alexander and {Zannese}, Marion and {Abergel}, Alain and {Bernard-Salas}, Jeronimo and {Boersma}, Christiaan and {Bron}, Emeric and {Chown}, Ryan and {Cuadrado}, Sara and {Dicken}, Daniel and {Elyajouri}, Meriem and {Fuente}, Asunci{\'o}n and {Gordon}, Karl D. and {Issa}, Lina and {Kannavou}, Olga and {Khan}, Baria and {Lacinbala}, Ozan and {Languignon}, David and {Le Gal}, Romane and {Maragkoudakis}, Alexandros and {Meshaka}, Raphael and {Okada}, Yoko and {Onaka}, Takashi and {Pasquini}, Sofia and {Pound}, Marc W. and {Robberto}, Massimo and {R{\"o}llig}, Markus and {Schefter}, Bethany and {Schirmer}, Thi{\'e}baut and {Sidhu}, Ameek and {Tabone}, Benoit and {Van De Putte}, Dries and {Vicente}, S{\'\i}lvia and {Wolfire}, Mark G.},
        title = "{Formation of the methyl cation by photochemistry in a protoplanetary disk}",
      journal = {\nat},
         year = 2023,
        month = sep,
       volume = {621},
       number = {7977},
        pages = {56-59},
          doi = {10.1038/s41586-023-06307-x},
archivePrefix = {arXiv},
       eprint = {2401.03296},
 primaryClass = {astro-ph.GA},
       adsurl = {https://ui.adsabs.harvard.edu/abs/2023Natur.621...56B}
}

@ARTICLE{Berne2014,
       author = {{Bern{\'e}}, O. and {Marcelino}, N. and {Cernicharo}, J.},
        title = "{IRAM 30 m Large Scale Survey of $^{12}$CO(2-1) and $^{13}$CO(2-1) Emission in the Orion Molecular Cloud}",
      journal = {\apj},
         year = 2014,
        month = nov,
       volume = {795},
       number = {1},
          eid = {13},
        pages = {13},
          doi = {10.1088/0004-637X/795/1/13},
archivePrefix = {arXiv},
       eprint = {1408.2999},
 primaryClass = {astro-ph.GA},
       adsurl = {https://ui.adsabs.harvard.edu/abs/2014ApJ...795...13B}
}

@ARTICLE{Nagy2015,
       author = {{Nagy}, Z. and {Ossenkopf}, V. and {Van der Tak}, F.~F.~S. and {Faure}, A. and {Makai}, Z. and {Bergin}, E.~A.},
        title = "{C$_{2}$H observations toward the Orion Bar}",
      journal = {\aap},
         year = 2015,
        month = jun,
       volume = {578},
          eid = {A124},
        pages = {A124},
          doi = {10.1051/0004-6361/201424220},
archivePrefix = {arXiv},
       eprint = {1405.3903},
 primaryClass = {astro-ph.GA},
       adsurl = {https://ui.adsabs.harvard.edu/abs/2015A&A...578A.124N}
}

@ARTICLE{Ford1975,
       author = {{Ford}, A. Lewis and {Docken}, Kate Kirby and {Dalgarno}, A.},
        title = "{Cross Sections for Photoionization of Vibrationally Excited Molecular Hydrogen}",
      journal = {\apj},
         year = 1975,
        month = sep,
       volume = {200},
        pages = {788-789},
          doi = {10.1086/153850},
       adsurl = {https://ui.adsabs.harvard.edu/abs/1975ApJ...200..788F}
}

\end{document}